\documentclass[journal=jpcafh,manuscript=article]{achemso}
\usepackage{amssymb,amsmath,cmbright}
\usepackage{bm} 
\usepackage{lmodern}
\usepackage[utf8]{inputenc}
\usepackage[version=4]{mhchem}
\usepackage{chemformula} 
\usepackage{chemfig}
\usepackage[T1]{fontenc} 
\usepackage{graphicx}
\usepackage{dcolumn}
\usepackage{siunitx}
\DeclareSIUnit\angstrom{\text{Å}}
\usepackage{multirow}
\usepackage[ruled]{algorithm2e}
\usepackage{csquotes}
\usepackage{hyperref}
\hypersetup{
    colorlinks=true,
    linkcolor=cyan,
    filecolor=magenta,      
    urlcolor=cyan,
    citecolor=cyan,
    }
\usepackage{ulem}

\DeclareMathVersion{sans}
\SetSymbolFont{operators}{sans}{OT1}{cmbr}{m}{n}
\SetSymbolFont{letters}{sans}{OML}{cmbr}{m}{it}
\SetSymbolFont{symbols}{sans}{OMS}{cmbr}{m}{n}
\SetMathAlphabet{\mathit}{sans}{OT1}{cmbr}{m}{sl}
\SetMathAlphabet{\mathbf}{sans}{OT1}{cmbr}{bx}{n}
\SetMathAlphabet{\mathtt}{sans}{OT1}{cmtl}{m}{n}
\SetSymbolFont{largesymbols}{sans}{OMX}{cmex}{m}{n}

\mathversion{sans}

\SectionNumbersOn

\newcommand\dd{\mathop{}\!\mathrm{d}} 
\newcommand{\pd}{\partial}
\newcommand{\rr}{\bm{r}}
\newcommand{\R}{\bm{R}}
\newcommand{\efphi}{\Phi(\rr;t,\R)}
\newcommand{\efchi}{\chi(\R;t)}
\newcommand{\vecpot}{\bm{A}}
\newcommand{\tdpes}{\varepsilon}

\title{On the interpretation of molecular photoexcitation with long and ultrashort laser pulses}

\author{Ji\v{r}\'{i} Jano\v{s}}
\email{jiri.janos@vscht.cz}
\affiliation[VSCHT]
{Department of Physical Chemistry, University of Chemistry and Technology, Technická 5, Prague 6, 166 28, Czech Republic}
\alsoaffiliation[UB]
{Centre for Computational Chemistry, School of Chemistry, University of Bristol, Bristol BS8 1TS, United Kingdom}

\author{Federica Agostini}
\affiliation[LCT]{Sorbonne Universit\'e, CNRS, LCT UMR 7616, Paris 75005, France}

\author{Petr Slav\'{i}\v{c}ek}
\affiliation[VSCHT]
{Department of Physical Chemistry, University of Chemistry and Technology, Technická 5, Prague 6, 166 28, Czech Republic}

\author{Basile F. E. Curchod}
\email{basile.curchod@bristol.ac.uk}
\affiliation[UB]
{Centre for Computational Chemistry, School of Chemistry, University of Bristol, Bristol BS8 1TS, United Kingdom}


\begin{document}

\begin{tocentry}
\end{tocentry}

\begin{abstract}

Photoexcitation is an inherent part of any photochemical or spectroscopic experiment, yet its impact on the excited-state dynamics is often overlooked. However, it is the excited molecular state, built upon photoexcitation and shaped by the characteristics of the light source, that determines the fate of the excited molecule and its subsequent photochemical reactions. In this work, we investigate how excited molecular states are built by different laser pulses, leveraging two representations of the molecular wave function: the Born--Huang expansion and the exact factorization. We explore the generation of two limiting cases: a stationary molecular state generated with a long laser pulse and an electronic wave packet created by an ultrashort (attosecond) laser pulse. The standard concepts of population transfer between electronic states, the resonance condition, or sudden vertical excitation, inherent to the Born--Huang representation and used by chemists to approximate the impact of photoexcitation on molecular systems, are brought to a new light by the exact factorization. 

\end{abstract}



\section{Introduction}

The Grotthuss--Draper law, the first law of photochemistry, states that only the light absorbed by a molecular system can cause a photochemical reaction.\cite{grotthuss1819firstlawphotochem,Draper1841firstlawphotochem} While this law may appear trivial nowadays, it highlights the key impact of photoexcitation on the outcomes of photochemical reactions, and perhaps more importantly, on the molecular state such a photoexcitation generates.

The properties of the light source employed to excite a molecule---typically a laser pulse for time-resolved experiments or a continuous-wave (CW) laser for energy-resolved studies---may tune the precise \textit{molecular state} generated upon photoexcitation.\cite{persico2018photochemistry,Suchan2018,janos2025selecting} This molecular state, characterized by a (often time-dependent) molecular wave function depending on both electronic and nuclear degrees of freedom, is therefore carved by the driving light source. When an ultrashort (attosecond) laser pulse with a broad energy spectrum is sent on a molecule, the resulting molecular state upon photoexcitation is formed by a coherent superposition of nuclear wave functions on multiple electronic states -- a molecular state coined 'an electronic wave packet' (where a wave packet characterizes a coherent superposition of eigenstates, here electronic ones). Such an electronic wave packet exhibits an actual non-stationary electronic dynamics that lies at the heart of attochemistry. Extending the laser pulse to a few tens of femtoseconds reduces the span of its energy spectrum, meaning that its application to a molecular system will usually target only a single electronic state, yet the molecular state formed can still be composed of a coherent superposition of eigenstates, this time of vibrational (nuclear) eigenstates. Such molecular states are called nuclear or vibrational wave packets. Vibrational wave packets feature nearly classical nuclear dynamics and have been the central quantity of interest in femtochemistry. Finally, long CW laser pulses have a well-defined frequency, which may target only a single vibrational state of a given electronic excited state, thereby forming a stationary molecular state called a vibronic state. It is often in the context of a photoexcitation triggered by a long CW laser that various well-known concepts in photochemistry are introduced, e.g., Franck-Condon factors, the resonance condition, or the Franck-Condon principle, which states that an electronic transition occurs, to a good approximation, without changes in the position of the nuclei of a molecule.\cite{michl,turro2009principles,persico2018photochemistry}  

This brief survey of photoexcitation hopefully highlights the key importance of the photoexcitation process in creating a new molecular state, whose properties may alter outcomes of photochemical reactions. Yet, it is central to realize that the perspective presented above and its vocabulary---electronic and vibrational eigenstates, wave packets---belong to a specific representation of the molecular wave function called Born--Huang (BH).\cite{Born1954} It is within this BH representation that a molecular wave function can be defined as a linear combination of static electronic states multiplied by time-dependent nuclear amplitudes, yielding a picture of nuclei evolving on static electronic potential energy surfaces and transferring between them.\cite{conicalintersection2004,worth2004beyond,Agostini2019} The BH representation is, however, not the only possible representation for the molecular wave function. An alternative and equally exact representation is the so-called exact factorization (EF), which represents the molecular state as a single product of a nuclear state and an electronic state, both time-dependent.\cite{abedi_exact_2010,Abedi2012} Within the single product representation offered by the EF, the vocabulary and concepts dear to photochemists---static electronic states and static potential energy surfaces guiding time-dependent nuclear wave functions---do not exist anymore.\cite{Agostini2019,Agostini2018b} 
As such, the EF depicts the building of excited molecular states by photoexcitation in a fundamentally different and truly unconventional picture.
Historically, shifting perspectives on established concepts in quantum mechanics often led to a deeper understanding and the development of novel practical approaches, as exemplified by the impact of path integrals or Bohmian dynamics. The perspective provided by the EF on the specific process of photoexcitation has remained unexplored to date. 

The central questions of this work are therefore: can switching to the EF representation of the molecular wave function bring a new picture and a new understanding of molecular photoexcitation? Does it question traditional textbook concepts that belong solely to the BH representation? Can the EF offer a new route for simulating molecular photoexcitation?

To address these questions, we developed a complete model for the photoexcitation of the hydroxyl radical, which allowed us to explore distinct photoexcitation scenarios: a long 100-fs laser pulse forming a stationary molecular state and a 1-fs laser pulse creating an electronic wave packet. The long 100-fs pulse depicts the gradual build-up of the stationary molecular state, whereas the short 1-fs laser pulse reveals the sudden formation of an electronic wave packet -- a scenario common in attochemistry.

Our results cast new light on our understanding of photoexcitation rooted in the BH representation, revealing a striking conceptual difference between the two representations of molecular states: (\textit{i}) a population transfer between stationary states in the BH expansion can be equally viewed as tunneling in the EF, (\textit{ii}) the implications of vertical excitation and the resonance condition, connected to the Franck--Condon principle, are highlighted by the EF and clarified by its formalism, and (\textit{iii}) the EF makes it possible to demonstrate pitfalls in simulating photoexcitation with a truncated basis (Hilbert space) that would emerge from a BH-based understanding of photoexcitation. 

The paper is organized as follows. Section~\ref{sec:theor_back} offers a pedagogical introduction to the theoretical background behind the BH and EF representations. The hydroxyl radical model and the methodology applied in this work are detailed in Section~\ref{sec:methods}. We then present our results and discuss their implications in Section~\ref{sec:results}, before drawing our conclusions and assessing their impact on the field of photochemistry in Section~\ref{sec:conclusions}.

\section{Theoretical background}
\label{sec:theor_back}

In this Section, we begin by introducing the concept of the molecular wave function, representing the molecular state discussed in the Introduction, and the corresponding time-dependent Schr\"odinger equation. We follow by discussing two different representations -- the BH expansion and the EF of the molecular wave function -- along with their equations of motion. Finally, we provide a connection between these two representations.

\subsection{Molecular wave function and Born--Oppenheimer approximation}

The object of interest throughout this Section is the molecular wave function, $\Psi$, which contains the complete information about the molecular state. The molecular wave function is, in the position representation, a complex function of the positions of all the particles composing a molecular system and of time $t$. Although all the particles are treated equally in quantum mechanics, we tend to formally distinguish them by marking the positions of lighter electrons with $\rr$ and heavier nuclei with $\R$. Using this notation, the molecular wave function is then written in the position representation as $\Psi(\rr,\R;t)$. Note that since the dependence on time in the molecular wave function is parametric in non-relativistic quantum mechanics, we separate it (and any other parametric dependence in any other wave function) by a semicolon from the remaining physical variables for wave functions.

The time evolution of $\Psi(\rr,\R;t)$ in a non-relativistic framework is governed by the time-dependent Schr\"odinger equation,
\begin{equation}
    i\hbar\frac{\pd}{\pd t} \Psi(\rr,\R;t) = \hat{H}(\rr,\R, t) \Psi(\rr,\R;t) \, ,
    \label{eq:tdse}
\end{equation}
where $\hat{H}(\rr,\R,t)$ is the molecular Hamiltonian defined as
\begin{align}
    \hat{H}(\rr,\R, t) &= \hat{T}_\mathrm{n}(\R) + \hat{T}_\mathrm{e}(\rr) + \hat{V}_\mathrm{nn}(\R) + \hat{V}_\mathrm{en}(\rr,\R) + \hat{V}_\mathrm{ee}(\rr) + \hat{V}_\mathrm{int}(\rr,\R,t) \, .
    \label{eq:molH}
\end{align}
The molecular Hamiltonian contains the following terms: kinetic energy operators for nuclei [$\hat{T}_\mathrm{n}(\R)$] and electrons [$\hat{T}_\mathrm{e}(\rr)$], Coulomb interactions between electrons and nuclei [$\hat{V}_\mathrm{nn}(\R)$, $\hat{V}_\mathrm{en}(\rr,\R)$, and $\hat{V}_\mathrm{ee}(\rr)$], and a time-dependent external interaction term [$\hat{V}_\mathrm{int}(\rr,\R,t)$]. 
The external interaction---here, the electromagnetic field of a laser pulse for our interest in photoexcitation---can affect both electrons and nuclei: $\hat{V}_\mathrm{int}(\rr,\R,t) = \hat{V}_\mathrm{int}^\mathrm{e}(\rr,t) + \hat{V}_\mathrm{int}^\mathrm{n}(\R,t)$. We will use this decomposition later when we discuss the equations of motion related to the EF.

To alleviate the complexity related to the full molecular wave function, a suitable representation decomposing $\Psi(\rr,\R;t)$ into quantities related to the electrons and the nuclei is desirable. The first and seminal step in this direction was made by Born and Oppenheimer,\cite{Born1927} who realized that electrons are much lighter than nuclei and can adapt to nuclear motion almost instantaneously under certain conditions. Formulating this assumption mathematically is equivalent to saying that, independently of the nuclear motion, electrons always remain adiabatically in a single  electronic state; in other words, the total molecular wave function can be written as a product of a time-dependent nuclear wave function $\chi(\R;t)$ and a time-independent electronic wave function $\Phi(\rr;\R)$,
\begin{equation}
    \Psi(\rr,\R;t) \approx \chi(\R;t)\Phi(\rr;\R) \,.
    \label{eq:boa}
\end{equation}
The electronic wave function depends parametrically on the nuclear positions $\R$, which gives its squared modulus a conditional probability nature, reflecting the assumption that the electrons can adapt to the nuclei instantaneously. The advantage of Eq.~\eqref{eq:boa} is that it separates the molecular wave function into two wave functions---one for nuclei and one for electrons---which can be analyzed (and visualized) separately. 

Where does the electronic wave function in Eq.~\eqref{eq:boa} come from? The electronic wave function can be taken as an eigenstate of the electronic subsystem, i.e., a solution of the time-independent electronic Schr\"odinger equation
\begin{align}
    \hat{H}_\mathrm{el}(\rr,\R)\Phi(\rr;\R) = \varepsilon(\R)\Phi(\rr;\R) \,,
    \label{eq:eltise}
\end{align}
where the electronic subsystem is defined by the electronic or Born--Oppenheimer Hamiltonian $\hat{H}_\mathrm{el}(\rr,\R)=\hat{T}_\mathrm{e}(\rr) + \hat{V}_\mathrm{nn}(\R) + \hat{V}_\mathrm{en}(\rr,\R) + \hat{V}_\mathrm{ee}(\rr)$. The quantity $\varepsilon(\R)$ is then the electronic energy, forming a potential energy surface for the nuclei when Eq.\eqref{eq:eltise} is solved for all possible nuclear configurations $\R$.~\cite{Curchod_FC2025} The Born-Oppenheimer molecular wave function is often expressed for the ground electronic state, i.e., $\Phi(\rr;\R) \equiv \Phi_0(\rr;\R)$ and $\varepsilon(\R) \equiv \varepsilon_0(\R)$. However, we can write this approximation to the molecular wave function for any electronic state, and we label these eigenstates by $J$: $\Phi_J(\rr;\R)$ and $\varepsilon_J(\R)$.  

The product form in Eq.~\eqref{eq:boa} is, however, only an approximation as such a simple product is only able to account for the full complexity of the molecular wave function under certain conditions. Hence, Eq.~\eqref{eq:boa} is known as the \textit{Born--Oppenheimer approximation} (BOA). The failures of BOA are well known: its simple product cannot describe nonadiabatic phenomena, that is, processes involving the coupling between nuclear and electronic motion. In addition, the simple BOA molecular wave function does not allow a rigorous description of photoexcitation between electronic states.  To create an in-principle exact representation for the molecular wave function, one needs to make a step beyond BOA.

\subsection{Born--Huang expansion of the molecular wave function}

The most straightforward way to correct the BOA wave function towards an exact representation relies on the property that the electronic Hamiltonian generates a set of orthonormal electronic eigenstates. The molecular wave function can be expanded in this basis with expansion coefficients $\chi(\R;t)$ depending on the remaining variables $\R$ and $t$. Mathematically speaking, we can suggest the following Ansatz,
\begin{equation}
    \Psi(\rr,\R;t) = \sum_{J} \chi_J(\R;t)\Phi_J(\rr;\R) \, ,
    \label{eq:bhansatz}
\end{equation}
which implies that we can form Born--Oppenheimer products \eqref{eq:boa} for each electronic state and sum them up. This Ansatz is known as the \textit{Born--Huang expansion} (BH) of the molecular wave function.\cite{Born1954} In the following, we summarize the main aspects of the BH expansion which are detailed in the following references: \citenum{prlj2025bestpracticesnonadiabaticmolecular,Baer2006, Tully2012, Agostini2019}.

As stated above, the electronic states are the eigenstates of the electronic Hamiltonian $\hat{H}_\mathrm{el}(\rr,\R)$; hence, they are obtained from the time-independent Schr\"odinger equation,
\begin{equation}
    \hat{H}_\mathrm{el}(\rr,\R)\Phi_J(\rr;\R) = \varepsilon_J(\R)\Phi_J(\rr;\R) \, ,
    \label{eq:eltise2}
\end{equation}
where $\varepsilon_J(\R)$ now represents an infinite set of electronic energies corresponding to different electronic states $J$. As such, electronic wave functions and electronic energies (that is, potential energy surfaces) do not depend on time explicitly and can be precomputed before any dynamical simulation. This stationary electronic basis is the advantage of the BH expansion as it allows us to plot the potential energy landscape of any molecule and use it to infer or even  possible dynamical scenarios.

Plugging the BH expansion [Eq.~\eqref{eq:bhansatz}] into the time-dependent Schr\"odinger equation [Eq.~\eqref{eq:tdse}] yields a set of coupled equations of motion for the nuclear amplitudes,
\begin{equation}
     i \hbar \frac{\pd}{\pd t} \chi_J(\R;t) = \sum_{K=1}^{\infty} \left[\left( \hat{T}_\mathrm{n}(\R) + \varepsilon_K(\R) \right) \delta_{JK} + \hat{D}_{JK}(\R) + \hat{V}_{\mathrm{int},JK}(\R,t)\right]\chi_K(\R;t) \, ,
     \label{eq:nucbhtdse}
\end{equation}
where $\hat{T}_\mathrm{n}(\R)=- \sum_{\alpha=1}^{N_{\text{n}}}\frac{\hbar^2\nabla_\alpha^2}{2M_\alpha}$ is the nuclear kinetic energy operator with $N_{\text{n}}$ being the number of nuclei and $\alpha$ their index, $\hat{D}_{JK}(\R)$ are the so-called nonadiabatic coupling terms,
\begin{align}
 \hat{D}_{JK}(\R) = & - \sum_{\alpha=1}^{N_{\text{n}}}\frac{\hbar^2}{2M_\alpha} \left[\langle\Phi_J(\R)|\nabla_\alpha^2 | \Phi_K(\R)\rangle_{\rr} +  2\langle{\Phi_J(\R)} | \nabla_\alpha \Phi_K(\R)\rangle_{\rr} \cdot\nabla_\alpha\right] \, ,\footnotemark
\end{align}\footnotetext{We adapt the bra-ket notation where $\langle\cdot\rangle_{\rr}$ indicates the inner product taken over the electronic degrees of freedom. Note that the kets $|\Phi_J(\R)\rangle$ or bras $\langle\Phi_J(\R)|$ carry no explicit dependence on the electronic coordinates $\rr$ as they denote electronic states, not coordinate-space functions. Their dependence on the nuclear coordinates $\R$ (and possibly on time $t$) is purely parametric.}

\noindent and $\hat{V}_{\mathrm{int},JK}(\R,t)$ represents a coupling term with the external potential,
\begin{equation}
    \hat{V}_{\mathrm{int},JK}(\R,t) = \langle \Phi_J(\R) | \hat{V}_\mathrm{int}(\R,t) | \Phi_K(\R) \rangle_{\bm{r}} \,.
\end{equation}
Solving these coupled equations requires prior knowledge of the electronic wave functions (for the nonadiabatic coupling and interaction terms) together with the electronic energies. Hence, the electronic time-independent Schr\"odinger equation [Eq.~\eqref{eq:eltise2}] must be solved before the nuclear propagation, in principle for any possible nuclear configuration. 

The dynamics of a molecular system based on Eq.~\eqref{eq:nucbhtdse} reads as follows: the nuclear amplitudes $\chi_J(\R;t)$ evolve on their respective potential energy surfaces $\varepsilon_J(\R)$, whilst being mutually coupled by the nonadiabatic couplings $\hat{D}_{JK}(\R)$ and the interaction term $\hat{V}_{\mathrm{int},JK}(\R,t)$. Due to these couplings, nuclear amplitude can be transferred from one electronic state to another and, as such, implicitly reflect the electron dynamics during nonadiabatic events (or photoexcitation). Hence, the electronic dynamics is encoded in the nuclear amplitudes as well as the nuclear dynamics within the BH representation. For example, a purely electronic excitation triggered by $\hat{V}_\mathrm{int}^\mathrm{e}(\bm{r},t)$ is described at the level of the nuclear amplitudes as a transfer of nuclear amplitude between different electronic states. 

Finally, we note that the full set of electronic states can also be transformed to a different basis, such as a diabatic basis.\cite{Yarkony2019} A diabatic basis might be more suitable for different kinds of applications, yet the conceptual picture remains the same.

\subsection{Exact factorization of the molecular wave function}

An alternative approach exists to offer an exact representation of the molecular wave function whilst retaining the appealing single-product form of BOA. The exact factorization  (EF) of the molecular wave function renounces the assumption that electrons instantaneously adapt to the nuclear motion by using a single, yet time-dependent electronic wave function which is still parametrically dependent on the nuclear positions. The single-product form of the molecular wave function offered by the EF reads
\begin{equation}
    \Psi(\rr,\R;t) = \efchi\efphi \,.
    \label{eq:efansatz}
\end{equation}
In this representation, it is no longer assumed that the electronic wave function is an eigenstate of the electronic Hamiltonian [as in BH, Eq.~\eqref{eq:eltise2}]. As such, the single-product form of the molecular wave function is not uniquely defined within the EF: if we were to multiply $\efchi$ by some nonzero factor and divide $\efphi$ by the same factor, we would obtain the same overall molecular wave function $\Psi(\rr,\R;t)$.\footnote{In contrast, the BH expansion is uniquely defined up to a phase (or even up to a sign for commonly used real electronic wave functions), if the electronic states are eigenstates of the electronic Hamiltonian. However, the total molecular wave function is invariant under a unitary transformation of the electronic basis, e.g. adiabatic to diabatic basis, which would transform as well the nuclear wavefunctions, leaving a certain freedom in the definition of the terms forming the BH expansion, similar to the gauge freedom of the EF (see Ref.~\citenum{conicalintersection2004} for additional details).} This ambiguity in the single-product Ansatz [Eq.~\eqref{eq:efansatz}] is naturally removed by the partial normalization condition (PNC), 
\begin{equation}
    \langle \Phi(t,\R)  | \Phi(t,\R) \rangle_{\rr}  = 1 \quad \forall \R, t \,,
\end{equation}
which allows us to identify the electronic wave function as a proper conditional amplitude, whose squared modulus yields a normalized conditional probability density.
The PNC makes the product in Eq.~\eqref{eq:efansatz} unique up to a gauge, which will be discussed later. Moreover, PNC gives $\efchi$ the meaning of a true nuclear wave function by reproducing the exact nuclear density: $|\efchi|^2=\int|\Psi(\rr,\R;t)|^2\dd\rr$. 
Formally, it is the combination of the Ansatz for the molecular wave function [Eq.~\eqref{eq:efansatz}] together with the PNC that strictly defines the \textit{exact factorization} of the molecular wave function. 

Thus, the EF uses a single wave function for the nuclei and a single wave function for the electrons, allowing for a simpler visualization compared to the large set of wave functions in the BH representation. Nevertheless, the fully time-dependent character of the EF does not allow to precompute its quantities before performing the dynamics. Consequently, the EF is less suitable for inferring possible dynamical scenarios beforehand, in this respect offering less predictive power than the BH representation.

The equations of motion in the EF are obtained by inserting the EF Ansatz into the time-dependent Schr\"odinger equation \eqref{eq:tdse} and imposing PNC, which leads to a set of two coupled time-dependent equations.\cite{Abedi2012,Agostini2018} The equation of motion for the electronic wave function reads 
\begin{equation}
    i \hbar \frac{\pd}{\pd t} \efphi = \left[ \hat{H}_\mathrm{el} (\R,t) +\hat{V}_\mathrm{int}^\mathrm{e}(\rr,t) + \hat{U}_\mathrm{en}[\chi,\Phi](\R,t) - \tdpes(\R,t) \right]\efphi \, ,
    \label{eq:efeomel}
\end{equation}
while the equation of motion for the nuclear wave function is 
\begin{equation}
    i \hbar \frac{\pd}{\pd t} \efchi = \left[ \sum_{\alpha=1}^{N_\mathrm{n}} \frac{[-i\hbar\nabla_\alpha + \vecpot_\alpha(\R,t)]^2}{2M_\alpha} + \hat{V}_\mathrm{int}^\mathrm{n}(\R,t) +\tdpes(\R,t) \right]\efchi \, .
    \label{eq:efeomnuc}
\end{equation}
Three new quantities were introduced in the previous equations, expressing the coupling between electrons and nuclei: the time-dependent vector potential $\vecpot(\R,t)$, the time-dependent potential energy surface $\tdpes(\R,t)$, and the electron-nuclear coupling operator $\hat{U}_\mathrm{en}[\chi,\Phi](\R,t)$. In the following, we will briefly discuss the interpretation of these three terms and the EF equations of motion. A detailed derivation and analysis of the EF equations can be found in the literature.\cite{Abedi2012, Agostini2018, Agostini2015, Ibele2024, Agostini2020}

The time-dependent vector potential (TDVP) for nucleus $\alpha$ is defined as 
\begin{equation}
    \vecpot_\alpha(\R,t) = - i \hbar \langle \Phi(t,\R)  | \nabla_\alpha \Phi(t,\R) \rangle_{\rr} 
    \label{eq:vecpot}
\end{equation}
and bears the character of a nuclear momentum field. In the EF, the total nuclear momentum is a sum of two contributions: from the nuclear wave function $\chi(\R;t)$ and from TDVP. Hence, TDVP captures the non-irrotational part of the nuclear momentum that is missing in the (curl-free component carried by the) nuclear wave function $\chi(\R;t)$, but is encoded in the electronic wave function $\Phi(\rr;t,\R)$ due to the single-product ansatz of the EF. 

The time-dependent potential energy surface (TDPES) appearing in both electronic and nuclear equations of motion is defined as
\begin{align}
    \tdpes(\R,t) = \langle \Phi(t,\R)  | \hat{H}_\mathrm{el} (\R,t) + \hat{V}_\mathrm{int}^\mathrm{e}(t) + \sum_{\alpha=1}^{N_\mathrm{n}}\frac{[-i\hbar\nabla_\alpha - \vecpot_\alpha(\R,t)]^2}{2M_\alpha} - i \hbar \frac{\pd}{\pd t} | \Phi(t,\R) \rangle_{\rr} 
    \label{eq:tdpes} \,.
\end{align}
The TDPES consists of four different terms and provides an exact time-dependent generalization of the Born--Oppenheimer potential energy surface. The meaning of the first two terms is intuitive: the electronic Hamiltonian term creates a mean-field potential energy surface similar to that obtained in Ehrenfest dynamics (but with the EF electronic wave function), while the electronic interaction term accounts for the response of the electrons to the external interaction. The meaning of the remaining two terms is more elusive, but they introduce the entanglement between electrons and nuclei missing from the mean-field Ehrenfest time-dependent potential energy surface.\cite{Agostini2019b}

Finally, the electron-nuclear coupling operator is given by
\begin{align}
    \hat{U}_\mathrm{en}[\chi,\Phi](\R,t) = \sum_{\alpha=1}^{N_\mathrm{n}}\frac{1}{M_\alpha}\Bigg[  \frac{[-i\hbar\nabla_\alpha - \vecpot_\alpha(\R,t)]^2}{2} + &  \left( \frac{-i\hbar\nabla_\alpha \efchi}{\efchi} + \vecpot_\alpha(\R,t) \right) \notag \\
     & \cdot \left(-i\hbar\nabla_\alpha
    - \vecpot_\alpha(\R,t)\right) \Bigg] \,.
\end{align}
Note that, unlike TDVP and TDPES which depend solely on the electronic wave function, the electron-nuclear coupling operator depends on both the nuclear and electronic wave functions. As such, this quantity brings a coupling to the nuclei in the equation of motion for the electronic wave function [Eq.~\eqref{eq:efeomel}], leading to its name: the electron-nuclear coupling operator. 
Analogously to TDVP, $\hat{U}_\mathrm{en}$ can be related to a contribution to the total nuclear kinetic energy that is missing in the nuclear wave function $\chi(\R;t)$ (see Ref.~\citenum{Abedi2012} for details). 

The three quantities described above---TDVP, TDPES, and the electron-nuclear coupling operator---are responsible for the coupling between electrons and nuclei within the EF framework. While the electron-nuclear coupling ties the electronic equation of motion [Eq.~\eqref{eq:efeomel}] to that for the nuclei [Eq.~\eqref{eq:efeomnuc}], TDVP and TDPES couple the nuclei [Eq.~\eqref{eq:efeomnuc}] to the electrons [Eq.~\eqref{eq:efeomel}]. Hence, the TDVP and TDPES can be useful interpretative tools in nonadiabatic quantum molecular dynamics as they unravel the electronic effects on nuclei.

Let us now inspect the electronic and nuclear equations of motion [Eq.~\eqref{eq:efeomel} and Eq.~\eqref{eq:efeomnuc}]. The electronic equation of motion is, due to its non-Hermitian Hamiltonian, a non-unitary Schr\"odinger-like equation, yet still preserving the norm.\cite{Tu2025} This equation of motion is also non-linear due to the electron-nuclear coupling operator $\hat{U}_\mathrm{en}$, making any numerical solution of this equation challenging.\cite{Gossel2019, Stetzler2026} The interaction of the electronic wave function with an external potential (e.g., the electric field of a laser pulse) appears in the electronic equation of motion both directly via the interaction operator $\hat{V}_\mathrm{int}^\mathrm{e}(\rr,t)$ and indirectly via the TDPES. Thus, the electronic response to the field is directly contained in the electronic equation of motion, unlike in the BH representation. 
On the other hand, the equation of motion for the nuclear wave function satisfies a standard time-dependent Schr\"odinger equation with a time-dependent scalar [$\tdpes(\R,t)$] and vector potentials [$\vecpot(\R,t)$], which couple the nuclei to the electrons.\cite{Tu2025} Hence, the nuclear wavefunction evolves on the support of the TDPES and under the influence of an additional momentum field provided by TDVP.  
The nuclei are also directly coupled to the external potential via the $\hat{V}_\mathrm{int}^\mathrm{n}(\R,t)$ operator. Note that the electronic response to the external potential manifests purely through TDPES and TDVP.

Finally, we comment on the gauge freedom inherent to the EF framework. The product form in Eq.~\eqref{eq:efansatz} is still not uniquely defined, even with the introduction of the PNC.\cite{Abedi2012} The product can be multiplied and divided by a phase factor $\mathrm{e}^{i\theta(\R,t)/\hbar}$ without altering the total wave function or any observables. 
The uniqueness of the product form of the molecular wave function within the EF is then guaranteed by fixing the gauge [either by directly selecting a functional form for $\theta(\R,t)$ or by indirectly imposing a condition on the gauge potentials]. 
Multiple gauges have been introduced in the EF.\cite{Agostini2020} The most common gauge sets the phase of $\chi(\R;t)$ to zero,\cite{Ibele2024} where all the nuclear momentum is thus contained in TDVP. Another common gauge, which is only possible in one-dimensional systems\cite{Agostini2018b}, sets TDVP to zero and contains all electronic effects in the TDPES.\cite{Agostini2015}

\subsubsection{Exact factorization for one nuclear degree of freedom}

In our specific case of a one-dimensional OH radical model, the nuclear equation \eqref{eq:efeomnuc} of the EF further simplifies. Firstly, the laser field polarization is selected such that the nuclear interaction term $\hat{V}_\mathrm{int}^\mathrm{n}(R,t)$ disappears and the laser field interacts directly only with the electronic subsystem (see Sec.~\ref{sec:ohradical}). Note that since the \ce{OH} radical is a one-dimensional system, we work with a single nuclear variable $R$. Secondly, we employ the gauge setting TDVP to zero, as mentioned above. Together, these two assumptions lead to a simpler nuclear equation of motion,
\begin{align}
    i \hbar \frac{\pd}{\pd t} \chi(R;t) = \left[ \hat{T}_\mathrm{n}(R) +\tdpes(R,t) \right]\chi(R;t) \,,
\end{align}
where the TDPES completely governs the nuclear dynamics. Hence, the TDPES constitutes a suitable tool for analyzing and interpreting nuclear dynamics. 

In multidimensional systems, the TDVP cannot be generally gauged away\cite{Agostini2018b} and must therefore be analyzed together with the TDPES. To highlight that our conclusions are not restricted only to one-dimensional systems by our selection of gauge but also extend to molecular systems, we provide results with a gauge setting the nuclear phase to zero as mentioned above in the Supporting Information.  

\subsection{Connection between the two representations of the molecular wave function}

We conclude by discussing the connections between the two representations of the molecular wave function. The BH and EF representations are connected by a one-to-one transformation,\cite{Schurger2025} which allows to transfer between the EF and BH quantities. For example, the total nuclear density is directly obtained from the EF nuclear wave function as $|\efchi|^2$, but it can also be reconstructed in the BH representation as $\sum_{J} |\chi_J(\R;t)|^2$, when the full electronic basis is considered. The comparison between the two representations proposed in this work is therefore not related to the accessibility of their respective quantities but to their interpretative power. We analyze each representation using the quantities naturally associated to it: the nuclear wave function $\efchi$, the TDVP $\vecpot(\R,t)$, and the TDPES $\tdpes(\R,t)$ in the EF, \textit{versus} the nuclear amplitudes $\chi_J(\R,t)$ and potential energy surfaces $\varepsilon_J(\R)$ in BH. In particular, although the total nuclear density can be obtained from the BH amplitudes, its evolution cannot be associated with a single BH potential energy surface. To obtain the effective scalar and vector potentials governing this nuclear density, one would also need to construct the TDPES and TDVP, which amounts to transforming the dynamics to the EF representation. Finally, we note that the perceived intuitiveness of the two representations may partly reflect the familiarity of chemists with the traditionally taught BH picture and its PESs, while an independent intuition for the EF framework remains less developed to date.

\section{Methods}
\label{sec:methods}

In this Section, we detail the methodology and the model system used in this work. We begin by defining our model system, inspired by the hydroxyl radical, and follow with a discussion of the two applied laser pulses. Finally, we provide details about our quantum dynamics calculations and the codes used.

\subsection{Model system inspired by the hydroxyl radical}
\label{sec:ohradical}

The model employed in this work is based on a two-state model of the hydroxyl (\ce{OH}) radical introduced by Korolkov and Paramonov in 1998.\cite{Korolkov1998} This model considers the lowest two diabatic electronic states of the \ce{OH} radical, $\mathrm{X}^2\Pi$ (ignoring its double degeneracy) and $\mathrm{A}^2\Sigma^+$, both being bound states.

Since we target complex photoexcitation processes, we extended this model by including two more diabatic states, $\mathrm{1}^2\Sigma^-$ and $1^2\Delta$, which both exhibit a dissociative character. Following Korolkov and Paramonov, we ignore the electronic states of higher spin multiplicity, as the spin-orbit coupling would facilitate a population transfer on a much longer (picosecond) timescale.\cite{Korolkov1998} It is this extended \ce{OH} radical model that we discuss in detail below, with additional validations and its relation to the original model of Korolkov and Paramonov detailed in the Supporting Information. We stress at this point that the present model is not \textit{per se} intended for quantitative modelling of the \ce{OH} radical, but it rather targets a qualitative exploration of photoexcitation processes of this molecule.  

\begin{figure}[ht!]
    \centering
    \includegraphics[width=0.95\textwidth]{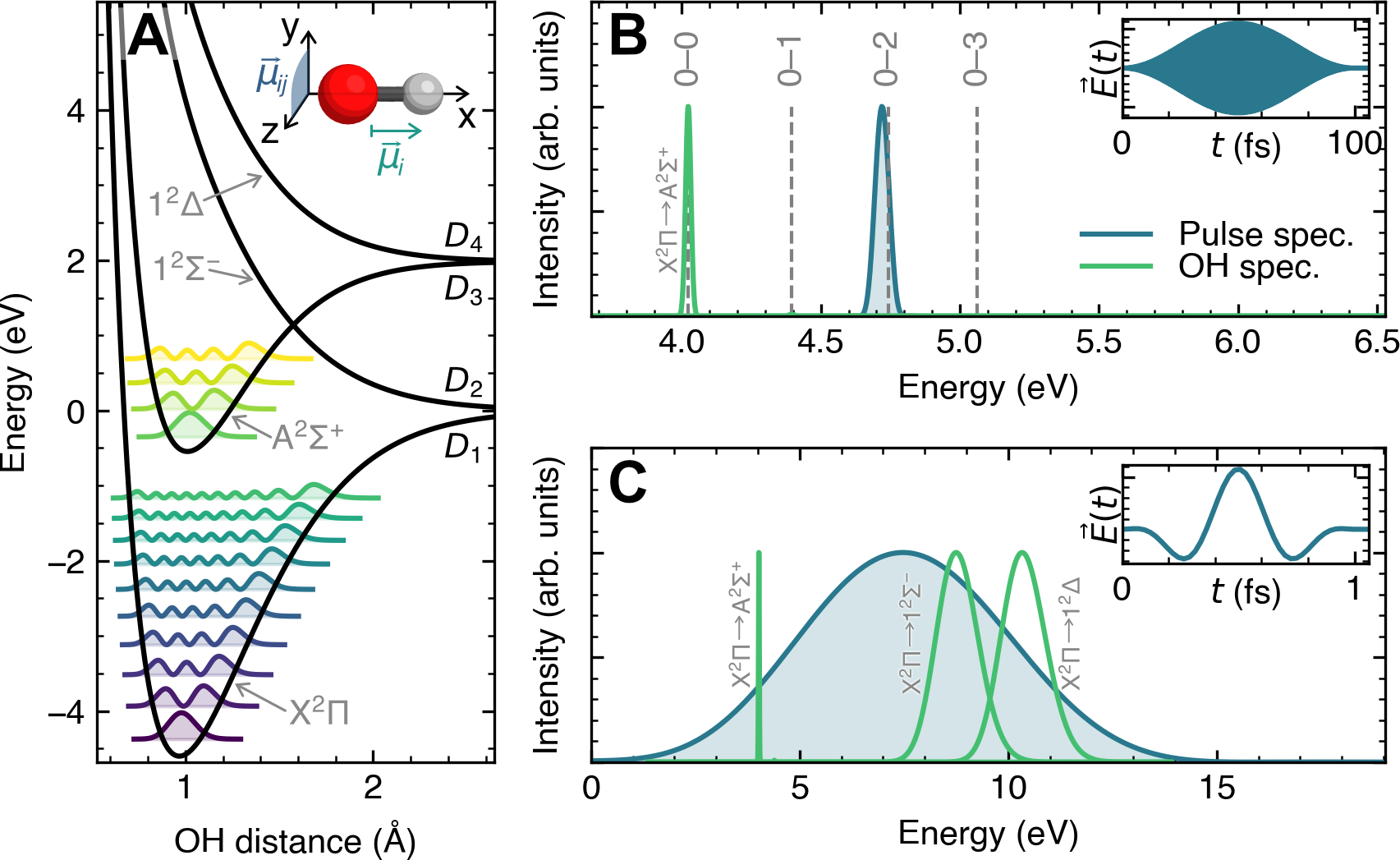}
    \caption{\textbf{A}: Potential energy curves of the \ce{OH} radical model with the relevant vibrational eigenstates of the lowest two diabatic states. The diabatic state labels ($\mathrm{X}^2\Pi$, $\mathrm{A}^2\Sigma^+$, $\mathrm{1}^2\Sigma^-$, $1^2\Delta$) are depicted together with the adiabatic state labels ($D_1$, $D_2$, $D_3$, $D_4$). The inset portrays the molecule and a schematic depiction of the dipole and transition dipole moments. \textbf{B}: Zoom on the region of the absorption spectrum of the \ce{OH} radical concerned with the $\mathrm{X}^2\Pi \rightarrow \mathrm{A}^2\Sigma^+$ electronic transition, depicted together with the spectral intensity of the 100-fs laser pulse used in our model (the pulse is shown in the inset). The grey dashed lines highlight the different vibronic transitions, demonstrating that only the 0--2 transition is resonant with the laser pulse employed in the present model. \textbf{C}: The normalized absorption spectra for the three excited states considered in the OH model are depicted together with the spectral intensity of the 1-fs laser pulse used in our model (the pulse is shown in the inset). The pulse spectral intensity targets all three excited states.}
    \label{fig:oh_ham}
\end{figure}

As stated above, our model features four diabatic electronic states: the bound $\mathrm{X}^2\Pi$ and $\mathrm{A}^2\Sigma^+$ states along with the dissociative $\mathrm{1}^2\Sigma^-$ and $1^2\Delta$ states (see Fig.~\ref{fig:oh_ham}A). Following Korolkov and Paramonov, we ignore the double degeneracy of both $\mathrm{X}^2\Pi$ and $1^2\Delta$ states. The diabatic potential energy curves are represented with an extended Rydberg function,
\begin{equation}
    V_{i}(R) = -D_i\left[1+ a_i(R-R_i) + b_i(R-R_i)^2 + c_i(R-R_i)^3\right]\exp\left[-\gamma_i(R-R_i)\right]+D_i^0 \,.
    \label{eq:extendedrydberg}
\end{equation}
The parameters for the diabatic potential energy curves of the $\mathrm{X}^2\Pi$ and $\mathrm{A}^2\Sigma^+$ states were reproduced from Korolkov and Paramonov,\cite{Korolkov1998} while those for the potential energy curves of the $\mathrm{1}^2\Sigma^-$ and $1^2\Delta$ states were fitted to electronic energies obtained with RMS-SA6-CASPT2(7e,5o) combined with a cc-pVTZ basis for O and a cc-pVDZ basis for H, calculated in OpenMolcas v22.06,\cite{openmolcas} see Tab.~\ref{tab:ohparam}. The diabatic couplings are zero between all electronic states due to symmetry. Hence, the adiabatic potential energy curves have the same shape as the diabatic ones; they are just reordered such that the adiabatic potential energy curves never cross. The adiabatic states are labelled $D_1$--$D_4$, highlighting the doublet spin multiplicity.

\begin{table}[ht!]
    \centering
    \renewcommand{\arraystretch}{1.2}
    \begin{tabular}{l c c c c c}
    \hline
    Parameter & $V_{\mathrm{X}^2\Pi}$ & $V_{\mathrm{A}^2\Sigma^+}$ & $V_{\mathrm{1}^2\Sigma^-}$ & $V_{1^2\Delta}$ & $\mu$ \\    
    \hline
    $a$ (a.u.$^{-1}$)         & $2.44875$ & $3.02658$ & $0.66099$ & $0.72427$ & $1.07015$ \\
    $b$ (a.u.$^{-2}$)         & $1.60999$ & $2.69337$ & $1.35253$ & $0.88319$ & $0.21032$ \\
    $c$ (a.u.$^{-3}$)         & $0.69090$ & $1.12086$ & $-1.19901\cdot 10^{-5}$ & $-0.21862$ & $-0.83399$ \\
    $\gamma$  (a.u.$^{-1}$)   & $2.48159$ & $3.06578$ & $2.31260$ & $2.24620$ & $1.60275$ \\
    $D$ (a.u.)                & $0.16905$ & $0.09297$ & $-0.24267$ & $-0.17649$ & $-0.28015$ \\
    $D^0$ (a.u.)              & $0.00000$ & $0.07294$ & $0.00000$ & $0.07294$ & $0.00334$ \\
    $R$ (a.u.)                & $1.84000$ & $1.92000$ & $1.54517$ & $1.71368$ & $0.98682$ \\
    \hline
    \end{tabular}
    \caption{Parameters of the extended Rydberg function [Eq.~\eqref{eq:extendedrydberg}] for the potential energy curves and the transition dipole moment ($\mu$) in the diabatic basis. Parameters for the $\mathrm{X}^2\Pi$ and $\mathrm{A}^2\Sigma^+$ states were reproduced from Ref.~\citenum{Korolkov1998}, only with a shift of the $D^0$ parameter by 75.49253 a.u. such that the first dissociation limit is at zero energy.}
    \label{tab:ohparam}
\end{table}

In the following, we will focus on the interaction between the \ce{OH} radical and two different laser pulses. The laser pulse interacts with the molecule through the dipole approximation: $-\hat{\vec{\mu}}(\rr,R)\cdot\vec{E}_0 E(t)$ where $\vec{\mu}(\rr,R)$ is the dipole moment operator, $\vec{E}_0$ is the polarization of the electric field, and $E(t)$ is the time-dependent magnitude of the electric field. The response of the molecule to the electric field is governed by the dipole matrix elements $\vec{\mu}_{IJ}(R) = \langle \Phi^\mathrm{(dia)}_I(R) |\hat{\mu} |\Phi^\mathrm{(dia)}_J(R)\rangle_{\rr}$, i.e., the dipole moments (diagonal elements) and transition dipole moments (off-diagonal elements). The interaction strength then depends on the projection of the (transition) dipole moments onto the polarization vector, i.e., $\vec{\mu}_{IJ}(R)\cdot\vec{E}_0$. In the case of the \ce{OH} radical, all dipole moments are aligned with the molecular axis and thus perpendicular to all the transition dipole moments (which are perpendicular to the molecular axis) -- see the inset in Fig.~\ref{fig:oh_ham}A. Thus, by defining the polarization of the laser pulse to be perpendicular to the molecular axis, the projection of the dipole moments onto the polarization vector of the pulse is zero, $\vec{\mu}_{II}(R)\cdot\vec{E}_0 = 0$, and the transition dipole moments are fully aligned with it, $\vec{\mu}_{IJ}(R)\cdot\vec{E}_0 = \mu_{IJ}(R)$. While this choice of alignment assumes preoriented molecules, e.g., by a strong IR field, it prevents vibrational transitions within a given electronic state and allows for purely electronic transitions (yet possibly with a vibrational component coupled with the electronic transition, as we shall see later). 

Finally, the transition dipole moments between excited states are zero due to symmetry, while our \textit{ab initio} calculations show that all transition dipole moments from the ground state share a similar magnitude and $R$-dependence. Hence, we simplify the model by approximating all the transition dipole moments from the ground electronic state to the same function:
\begin{equation}
    \vec{\mu}_{\mathrm{X}^2\Pi ,J}(R)\cdot\vec{E}_0 = \mu(R)\, , \quad\forall J \,.
\end{equation}
The transition dipole moment $\mu(R)$ was also fitted to the extended Rydberg function using the data from Korolkov and Paramonov\cite{Korolkov1998}; the fitting parameters are presented in Tab.~\ref{tab:ohparam}. Both calculated and fitted  potential energy curves and transition dipole moments are presented in Fig.~S1 in the Supporting Information together with an extended discussion of the model. 

Bringing together all the aforementioned elements, we obtain the final diabatic electronic Hamiltonian for our four-state model of the \ce{OH} radical under the influence of an aligned laser pulse,
\begin{equation}
    \mathbb{H}^\mathrm{(dia)}  = 
    \begin{pmatrix}
        V_{\mathrm{X}^2\Pi} & 0 & 0 & 0 \\
        0 & V_{\mathrm{A}^2\Sigma^+} & 0 & 0 \\
        0 & 0 & V_{\mathrm{1}^2\Sigma^-} & 0 \\
        0 & 0 & 0 & V_{1^2\Delta} \\
    \end{pmatrix}
    - 
    \begin{pmatrix}
        0 & \mu & \mu & \mu \\
        \mu & 0 & 0 & 0 \\
        \mu & 0 & 0 & 0 \\
        \mu & 0 & 0 & 0 \\
    \end{pmatrix}  E(t) \,,
\end{equation}
which is used for the quantum dynamics simulations described in Section~\ref{sec:qdmethods}. The diabatic wave functions and potential energy curves are transformed to the adiabatic picture by diagonalization of $\mathbb{H}^\mathrm{(dia)}$. From now on, we will prioritize the adiabatic basis in our discussion and use the adiabatic states labels $D_1$--$D_4$.

\subsection{100-fs and 1-fs laser pulses}
\label{sec:laserpulses}

The laser pulses employed in this work have the form of cosine oscillations modulated with a square sine envelope,
\begin{equation}
    \Vec{E}(t) = E_0 \sin^2\left(\frac{\pi t}{T}\right)\cos\left(\omega (t - T/2)\right) \quad  \textrm{for} \quad t  \in [0, T] \, ,
\end{equation}
where $E_0$ is the strength of the interaction, $T$ is the duration of the pulse, and $\omega$ is the frequency of the electric field oscillations.\cite{Korolkov1998} 
Although laser pulses defined via the electric field $\Vec{E}$ instead of a vector potential are not generally in line with Maxwell's equations, especially for short laser pulses, we carefully checked that our pulses do not violate Maxwell's equations, following Ref.~\citenum{Madsen2002} and SI of Ref.~\citenum{Janos2024}.

We applied two different laser pulses in our work: a 100-fs pulse and a 1-fs pulse. The 100-fs pulse was motivated by Korolkov and Paramonov, who tuned such pulses to achieve 100\% population transfer from the ground state ($D_1$) to selected vibrational states of the first excited electronic state $D_2$. In this work, we reoptimized the pulse parameters to trigger a full population transfer to the third (second excited) vibrational state in the $D_2$ electronic state by keeping the duration of the pulse but varying the intensity $E_0$ and frequency $\omega$. Our tuned laser pulse parameters are $T=100\,\text{fs}$, $E_0=0.0704\,\text{a.u.}$, and $\omega = 0.17365\,\text{a.u.} = 4.7253\,\text{eV}$. The pulse and its spectral intensity are depicted in Fig.~\ref{fig:oh_ham}B together with the normalized absorption spectrum of the $D_2$ ($\mathrm{A}^2\Sigma^+$) state. Note that the pulse spectrum is in resonance solely with the corresponding $|D_1,v=0\rangle \rightarrow|D_2, v=2\rangle$ vibronic transition, and that the other vibronic transitions are far from resonance.

The second pulse we chose has a duration of 1 fs and mimics a typical pulse used in the context of attochemistry. The parameters of this laser pulse are $T=1\,\text{fs}$, $E_0=0.5\,\text{a.u.}$, and $\omega = 0.27655\,\text{a.u.} = 7.525\,\text{eV}$. The pulse and its spectral intensity are depicted in Fig.~\ref{fig:oh_ham}C together with the normalized absorption spectrum of all the excited states. This ultrashort pulse targets all three excited states due to its sufficiently broad spectrum, and a coherent superposition of excited electronic states---an electronic wave packet---is expected to be formed by the pulse, as is typical in attochemistry experiments.

\subsection{Quantum dynamics of laser-excited \ce{OH} in the BH and EF representations}
\label{sec:qdmethods}

The quantum dynamics of the \ce{OH} radical under the influence of the 100-fs and 1-fs laser pulses were carried out in the BH representation using a diabatic basis with our in-house code QDyn v1.0.\cite{qdyn} The nuclear propagation employed a split-operator technique using the Fourier method for evaluating the kinetic energy operator. The time step was set to 0.1 a.t.u.\cite{Korolkov1998} The distance $R$ between the atoms of the \ce{OH} molecule ranged from 0.5 to 8.0 a.u., ensuring that the nuclear wave packets never reach the boundary of the grid in the present simulations, and was discretised into 2048 grid points. Convergence with respect to both the time step and the grid size was verified. The diabatic wave functions and potential energy curves were transformed to the adiabatic basis by diagonalizing the diabatic Hamiltonian after propagation.

The EF quantities were not directly propagated using the EF equations of motion, as this is unnecessarily laborious task. Instead, they were extracted from the diabatic BH wave functions.\cite{Ibele2022, Agostini2015, Curchod2016, Agostini2018b} Note that obtaining EF quantities from the BH wave function is equivalent to their direct propagation, as there is a one-to-one transformation between BH expansion and the EF.\cite{Schurger2025} This approach serves the purpose of demonstrating the two perspectives of photoexcitation since we are not interested in the EF propagation itself but rather in the behaviour of the quantities unique to the EF formalism. The EF quantities were obtained under a gauge in which the TDVP is set to zero, i.e., $A(R)=0$. Results with a gauge, in which the nuclear phase is set to zero, are provided in the Supporting Information. The EF quantities were directly obtained from the QDyn v1.0 code.\cite{qdyn}

Furthermore, we performed quantum dynamics simulations in the BH representation using a truncated basis (Hilbert space), i.e., by expanding the full molecular wave function as a linear combination of only selected vibronic states:
\begin{equation}
    \Psi(\rr, R; t) = \sum_{K=1}^{N_{\text{el}}} \sum_{v=0}^{N_{\text{vib}}} c_{K,v}(t)\chi_{K,v}(R)\Phi_K(\rr;R) \, ,
    \label{eq:decompstate}
\end{equation}
where $\chi_{K,v}(R)$ is the $v^\text{th}$ vibrational eigenstate of the $K^\text{th}$ electronic state $\Phi_K(\rr;R)$, and $c_{K,v}(t)$ are the time-dependent expansion coefficients of the vibronic states $\chi_{K,v}(R)\Phi_K(\rr;R)$. In this setup, only the expansion coefficients are time-dependent and propagated in the dynamics.
The purpose of these calculations was to inspect the role of specific vibronic states in the dynamics of \ce{OH} and assess the effect of basis set truncation on the outcomes of simulations. 
The vibrational eigenfunctions were obtained by imaginary-time quantum dynamics, as implemented in QDyn v1.0.\cite{qdyn} The Hamiltonian matrix elements between the vibronic states are provided in the Supporting Information.
The quantum dynamics with a truncated basis were carried out with a 0.1~a.t.u. time step in the cQDyn v1.0 code.\cite{cqdyn}.

\section{Results and Discussion}
\label{sec:results}

We begin our analysis with the formation of a stationary molecular state upon photoexcitation with a 100-fs laser pulse. The BH and EF pictures of photoexcitation are contrasted, which allows us to offer a different interpretation of a stationary state photoexcitation than the one offered in monographs and textbooks.\cite{Wardle2009, persico2018photochemistry, SchatzRatner2002_QMInChemistry, Tannor2007_IntroQMTimeDependent, AtkinsFriedman2010_MolecularQM, Engel2018_QuantumChemistrySpectroscopy, turro2009principles, michl} Follows a deeper investigation of the BH picture and the effect of basis set truncation, discussing the standard textbook concepts of the resonance condition and vertical excitation. We then discuss the impact of a short, 1-fs laser pulse on the excited molecular state and the nuclear dynamics driven by a coherent superposition of excited electronic states. 

\subsection{Photoexcitation to a stationary molecular state}

Let us start by discussing the photoexcitation of OH by the 100-fs laser pulse described in Sec.~\ref{sec:laserpulses}. The laser pulse has a very narrow spectral intensity profile targeting a single vibronic eigenstate of the hydroxyl radical (see Fig.~\ref{fig:oh_ham}B), resulting in the formation of a stationary molecular state (not a wave packet) after the pulse interaction ceases. Thus, the only dynamics triggered by this laser pulse should take place during the photoexcitation process, connecting the initial and final stationary states. The expected steps of this process are: OH in its stationary ground electronic and vibrational state (<0 fs) $\rightarrow$ coupled electron-nuclear dynamics during the pulse interaction (0--100 fs) $\rightarrow$ OH in a stationary vibrational state in the first excited electronic state (>100 fs). As such, the dynamics can be attributed purely to the formation of the excited molecular state, which is central to this paper.

\begin{figure}[ht!]
    \centering
    \includegraphics[width=1.0\linewidth]{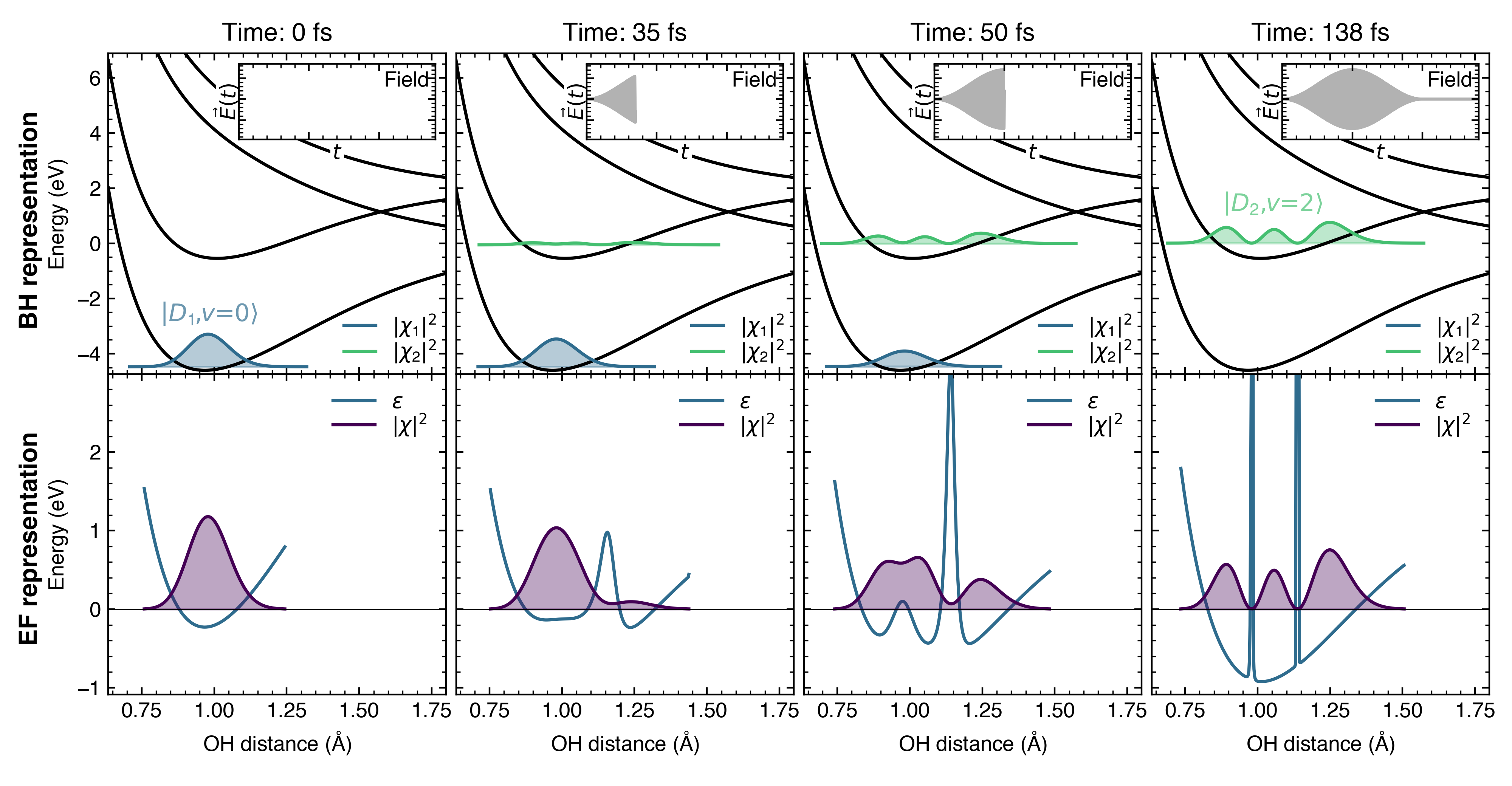}
    \caption{Snapshots of the OH photoexcitation with the 100-fs laser pulse in the BH (top row) and EF (bottom row) representations at four different times: 0, 35, 50, and 138 fs. The initial molecular state before the laser pulse interaction is observed at 0 fs (first column), the final stationary molecular state is observed for $t > 100$ fs -- the snapshot at 138 fs (last column) in this figure. Top row: the BH nuclear densities for the adiabatic states ($|\chi_I|^2$) and their corresponding adiabatic potential energy curves (black) are represented. Bottom row: the total nuclear density ($|\chi|^2$) and the TDPES ($\tdpes$) are depicted. The insets contain the electric-field component of the laser pulse [$\vec{E}(t)$] up to the given time of the snapshot. An animated version of the plot is available using the following \href{https://github.com/JanosJiri/Photoexcitation-From-Different-Perspectives/blob/main/100-fs pulse/BH_vs_EF_A0.gif}{link}.\cite{movierepo}}
    \label{fig:stata0}
\end{figure}

We now investigate the photoexcitation mechanism from the BH and EF perspectives, focusing on the dynamics during the laser pulse, which connects the two molecular stationary states. Snapshots of the dynamics in both representations are presented in Fig.~\ref{fig:stata0} (an animation of the process is available using the following \href{https://github.com/JanosJiri/Photoexcitation-From-Different-Perspectives/blob/main/100-fs pulse/BH_vs_EF_A0.gif}{link} for full inspection\cite{movierepo}).

We shall start with the BH picture as it constitutes the typical representation of photoexcitation taught to chemists (top row of Fig.~\ref{fig:stata0}). 
The 100-fs laser pulse fully depopulates the initial stationary molecular state (ground electronic and vibrational state $|D_1, v=0\rangle$), transferring the whole population to the third vibrational state of the first excited electronic state ($|D_2, v=2\rangle$). The final state remains stationary after the end of the pulse (>100 fs). The exclusive formation of the $|D_2, v=2\rangle$ molecular state strictly obeys the resonance condition, as the laser pulse spectrum overlaps only with the $|D_1,v=0\rangle\rightarrow|D_2,v=2\rangle$ transition (see Fig.~\ref{fig:oh_ham}B). During the photoexcitation, the molecular process appears simple: the ground molecular state gradually depopulates while the $|D_2, v=2\rangle$ state gains population. The BH nuclear densities for the two electronic states considered here ($|\chi_1|^2$ and $|\chi_2|^2$) seemingly do not exhibit any dynamics except for gaining/losing amplitude in the full scale plot depicted in the top row of Fig.~\ref{fig:stata0}. Hence, the photoexcitation appears to be a simple transfer of nuclear amplitude between two vibrational eigenstates in different electronic states: a picture commonly introduced in photochemistry textbooks to describe a stationary photoexcitation process. We shall see soon (Sec.~\ref{sec:diffintphotoexBH}) that this picture is actually more complex when examined in greater detail.

Moving to the EF, we now scrutinize the same process through the lens of the total nuclear density $|\chi(\R;t)|^2$ and the TDPES $\tdpes(\R,t)$ (bottom row of Fig.~\ref{fig:stata0}). 
The initial TDPES matches the $D_1$ potential energy curve, and $|\chi|^2$ corresponds to the ground-state nuclear density. Once the laser pulse arrives, the TDPES starts evolving and develops a new minimum around 1.25~Å, see the snapshot of 35~fs in Fig.~\ref{fig:stata0}. This new minimum, separated from the original minimum (at around 1~Å) by an energy barrier, draws the nuclear density towards longer bond lengths. We can identify this process as nuclear tunneling through the electronic TDPES, with the barrier height dictating the tunneling rate. To verify that the observed process can be associated to tunneling, we calculated the expectation value of the nuclear Hamiltonian, i.e., the total nuclear energy, defined in the EF as $E_\chi(t) = \big\langle \chi(t)\big|\sum_\alpha\frac{[-i\hbar\nabla_\alpha + \vecpot_\alpha(\R,t)]^2}{2M_\alpha} +\varepsilon(\R,t) \big| \chi(t) \big\rangle_{\R}$. The value of $E_\chi(t)$ stays at zero throughout the simulation, with slight oscillations of $\pm 0.06$~eV during the pulse. Therefore, the regions where the TDPES rises above zero should be considered as classically forbidden, and the nuclear flux through these regions corresponds to tunneling.

After 50 fs of irradiation, at the maximum of the laser pulse, enough nuclear density has reached the new minimum around 1.25~Å, and the increasing barrier height starts effectively reducing the tunneling and stabilizing localization at 1.25~Å. Moreover, a new minimum has emerged around 1.1~Å by that time, giving rise to another energy barrier which begins to further split the nuclear density.\footnote{We recommend watching the animations as they better depict the development of the barrier than the snapshots presented here.} Once the laser pulse interaction is over, both the TDPES and the nuclear density remain stationary. The TDPES mimics the $D_2$ potential energy curve except for the two singularities that developed from the energy barriers. These singularities create the nodes in the nuclear density and effectively block any tunneling. Note that quantitatively the same results are achieved in the gauge in which the nuclear phase is equal to zero (a gauge common in multidimensional molecular systems), suggesting the validity of the presented results across gauges, see the Supporting Information. 

Four clear and major observations emerge from comparing the BH and EF pictures of the photoexcitation presented above:
 
\textit{Observation 1}: The EF clearly distinguishes electronic and nuclear dynamics, as these are captured separately by $\efphi$ and $\efchi$. Since the laser field interacts only with the electrons in the present case, the nuclear wave function is unaffected in the early stages of the photoexcitation and starts evolving only later in response to the excited electrons. Hence, the EF helps us to easily identify that the laser field affects the nuclei only indirectly through their interaction with the electrons via the TDPES. In contrast, the BH nuclear amplitudes, $\chi_I(\R,t)$, encompass both nuclear and electronic dynamics, which makes it challenging to distinguish between the electronic and the nuclear contribution to the evolution. Since the nuclear amplitudes evolve in the electronic states and transfer between them, a more detailed analysis is necessary to separate electronic and nuclear contributions and draw the conclusions above.
    
\textit{Observation 2}: Both representations use different concepts to capture the photoexcitation: the BH representation features amplitude transfer between electronic states, while the EF representation employs barriers on the TDPES and tunneling through them to capture the same effects.

\textit{Observation 3}: The final nuclear density ($|\chi|^2$ in the EF and $|\chi_2|^2$ in BH) has two nodes irrespective of the representation; however, these nodes have a different origin. While the nodes in the BH representation come naturally from the orthogonality of bound vibrational eigenfunctions~\cite{CouranHilbert_book_1953,teschl2014mathematical}, the nodes in the EF representation are enforced by the singularities in the TDPES. This observation is, however, not directly related to the photoexcitation process \textit{per se}.\footnote{The appearance of nodes in the nuclear wave function is not in disagreement with the extensive literature devoted to nodeless vibrational wave functions in the context of the EF, as for instance Refs.~\citenum{Hunter_IJQC1980, Hunter_IJQC1981, Jonas_CPL2017, Schild_PRR2021}. Specifically, when applying the EF to stationary molecular eigenstates, the nuclear time-dependent Schr\"odinger equation~\eqref{eq:efeomnuc} is replaced by a stationary Schr\"odinger equation featuring the exact potential energy surface -- and the exact vector potential in more than one dimension. The lowest energy solution, yielding a nodeless wave function in the absence of symmetries, provides the vibrational component of the EF.~\cite{Jonas_CPL2017, Schild_PRR2021} While nodeless, such wave functions are reminiscent of the result shown in Fig.~\ref{fig:stata0} at 138~fs since they are associated with `spiky', yet non-singular, exact potential energy surfaces. By contrast, when the BOA limit is approached, the spiky potential develops singularities~\cite{Schild_PRR2021} as is the case at the end of the photoexcitation process with the 100-fs laser pulse.}
        
\textit{Observation 4}: The EF highlights the rich nuclear dynamics through the total nuclear density and the electronic dynamics through the TDPES during the photoexcitation process. In plain terms, both TDPES and the total nuclear density visibly evolve during the dynamics: the TDPES creating new minima and developing barriers, the nuclear density tunneling through these barriers and occupying the minima. Contrarily, the same dynamics are less apparent at first sight in the BH representation, seemingly featuring only a population transfer. 

The Observations above highlight the conceptual difference between the BH and EF representations and opportunities stemming from a change of perspective. Ensuing the last point, we now further analyze the BH quantities and extract the full dynamical picture seen in the EF, disentangling the nuclear and electronic contributions to the dynamics. 

\subsubsection{A different interpretation of photoexcitation in the Born--Huang picture}
\label{sec:diffintphotoexBH}

First, let us recall that the pulse spectral intensity effectively encompasses only the transition to the $|D_2,v=2\rangle$ state; all other states lie far off resonance with the laser pulse employed. In particular, the spectral intensity for the excitation to $|D_2,v=1\rangle$ is only $0.0000007\%$ of that to $|D_2,v=2\rangle$. Following first-order time-dependent perturbation theory---the common textbook approach to linear spectroscopy with long or continuous pulses---and the ensuing resonance condition, the molecular wave function can be expanded only in the basis of these two resonant vibronic states, $\chi_{D_1,v=0}(R)\,\Phi_{D_1}(\mathbf{r};R)$ and $\chi_{D_2,v=2}(R)\,\Phi_{D_2}(\mathbf{r};R)$, where $\chi_{D_1,v=0}(R)$ and $\chi_{D_2,v=2}(R)$ are the nuclear eigenfunctions for $v=0$ and $v=2$ in their respective electronic state. 
This resonance-based argument is central to standard spectroscopic reasoning and is consistent with the simulated dynamics in Fig.~\ref{fig:stata0}, which end in a pure $|D_2,v=2\rangle$ state. Hence, if the two-state expansion were an adequate approximation, we would expect an essentially complete population transfer into $|D_2,v=2\rangle$.

\begin{figure}[ht!]
    \centering
    \includegraphics[width=0.65\linewidth]{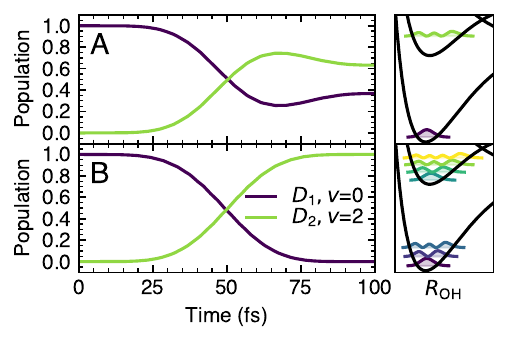}
    \caption{Populations of the $|D_1,v=0\rangle$ (purple) and $|D_2,v=2\rangle$ (green) states for simulations with a truncated expansion of the vibronic states: \textbf{A} two vibronic states, \textbf{B} seven vibronic states. The vibrational eigenstates used in the expansion are depicted in the right panels.}
    \label{fig:trunkHspace}
\end{figure}

To test this two-state hypothesis, we can now expand the time-dependent molecular wave function in the basis of the two vibronic states described above (following Eq.~\eqref{eq:decompstate}), 
\begin{equation}
    \Psi(\mathbf{r},R;t) =
    c_1(t)\chi_{D_1,v=0}(R)\,\Phi_{D_1}(\mathbf{r};R)
    + c_2(t)\chi_{D_2,v=2}(R)\,\Phi_{D_2}(\mathbf{r};R) \,.
\end{equation}
Before the laser pulse, the molecular wave function is described by $c_1(t)=1$ and $c_2(t)=0$, that is, a stationary OH molecule in its ground vibrational and electronic state. We can then propagate this approximate molecular wave function under the effect of the 100-fs laser pulse and depict the populations ($|c_1(t)|^2$ and $|c_2(t)|^2$) of the two resonant vibronic states, $|D_1,v=0\rangle$ and $|D_2,v=2\rangle$. The resulting dynamics, depicted in Fig.~\ref{fig:trunkHspace}A, show that this two-state picture is evidently invalid. The two-state approximation does not capture the full population transfer to the $|D_2,v=2\rangle$ state, with the final population being only 0.63. Thus, the other off-resonant vibronic states must participate in the dynamics, although they are not apparent at first sight in Fig.~\ref{fig:stata0}. Extending our truncated expansion to the three lowest vibrational states of $D_1$ and four lowest vibrational states of $D_2$ allows us to capture the full population transfer (Fig.~\ref{fig:trunkHspace}B) in agreement with the full quantum-dynamics simulation presented above. The addition of any further vibronic states does not alter the resulting dynamics. This set of simulations highlights the dramatic role played by the neighboring vibronic states in the description of a stationary photoexcitation, even though they are off-resonant. Simulations with various other combinations of vibronic states in the expansion are provided in the Supporting Information.

\begin{figure}[ht!]
    \centering
    \includegraphics[width=0.6\linewidth]{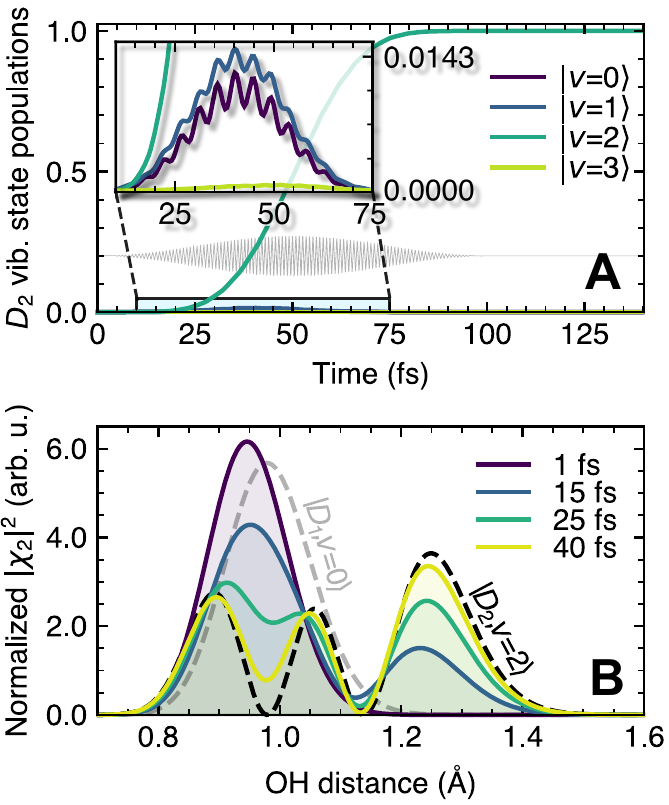}
    \caption{\textbf{A}: Populations of the vibrational states belonging to electronic state $D_2$ during the photoexcitation process. The legend uses a simplified notation where $|D_2,v=n\rangle$ is denoted as $|v=n\rangle$. The inset focuses on the off-resonant vibrational states between 10 and 75~fs.
    \textbf{B}: Normalized (time-dependent) nuclear density on $D_2$ ($|\chi_2|^2/\langle\chi_2|\chi_2\rangle$) for selected times of 1, 15, 25, and 40~fs. The dashed lines depict the normalized (stationary) nuclear densities of the initial $|D_1,v=0\rangle$ state (grey) and final $|D_2,v=2\rangle$ state (black).}
    \label{fig:vibdecomp}
\end{figure}

To understand the impact of neighbouring vibronic states in the photoexcitation dynamics, let us return to the original quantum dynamics simulation in the BH representation (Fig.~\ref{fig:stata0}A) and decompose the $D_2$ nuclear amplitude $\chi_2(R;t)$ into the vibrational eigenfunctions for this electronic state, see Fig.~\ref{fig:vibdecomp}A. The second vibrational state, $|D_2,v=2\rangle$, dominates the decomposition; the remaining vibrational states are barely visible in the graph on a full scale. The maximum population among the off-resonant states is reached by $|D_2,v=1\rangle$, peaking to 1.5\% at 40~fs, closely followed by the population of $|D_2,v=0\rangle$. A non-negligible population is also acquired by the $|D_2,v=3\rangle$ state. Despite such low populations, these intermediate vibrational states account for the large discrepancy observed in Fig.~\ref{fig:trunkHspace} and act as mediators for the full population transfer, whereas omitting these vibronic states gives a final population of only 63\% .

While the crucial role of the off-resonant states may seem somewhat surprising from the perspective of the resonance condition, their necessity becomes clearer by focusing on the excited-state density $|\chi_2|^2$, see Fig.~\ref{fig:vibdecomp}B. Note that the initial $|D_1,v=0\rangle$ state and the final $|D_2,v=2\rangle$ state densities do not fully overlap in configuration space: the maximum of the $|D_2,v=2\rangle$ state density lies in a region around 1.25~Å where the initial $|D_1,v=0\rangle$ density decays effectively to zero. Since nuclei cannot \textquote{jump} through space, the nuclear density must flow smoothly and seamlessly to connect the initial and final density, as clearly depicted by the EF simulations in Fig.~\ref{fig:stata0}. The normalized nuclear density in $D_2$ at 1~fs (dark purple curve in Fig.~\ref{fig:vibdecomp}B), i.e., right after the pulse is initiated, resembles the initial nuclear density for $|D_1,v=0\rangle$ (dashed grey curve in Fig.~\ref{fig:vibdecomp}B); the shift to shorter bond lengths is a consequence of the linear transition dipole moment. We note that describing the excited nuclear wave function following photoexcitation as the ground-state nuclear wave function multiplied by the (position-dependent) transition dipole moment is known as the `sudden/vertical approximation' in physical and theoretical chemistry.\cite{prlj2025bestpracticesnonadiabaticmolecular} As time progresses, the nuclear density in $D_2$ gradually evolves towards longer bond lengths and populates the region around 1.25~Å, also forming a node at 0.98~Å. The off-resonant vibrational states are necessary to facilitate this smooth flow of nuclear density, and their population increases as seen in the inset of Fig.~\ref{fig:vibdecomp}A. At 40~fs, the nuclear density in $D_2$ (yellow curve in Fig.~\ref{fig:vibdecomp}B)  almost reaches the shape of the final $|D_2,v=2\rangle$ state (dashed black curve in Fig.~\ref{fig:vibdecomp}B). Thus, the off-resonant vibrational states are no longer necessary to accommodate the nuclear density, and their populations decline. The off-resonant vibrational states are, therefore, indispensable intermediates facilitating a smooth transfer of nuclear density from the sudden/vertical excitation to the nuclear density of the final state. Without the off-resonant vibrational states, the nuclear density would be forced to suddenly jump in configuration space.\footnote{Note that for the two-state approximate model in Fig.~\ref{fig:trunkHspace}A, the nuclear density in $D_2$ depicted in Fig.~\ref{fig:vibdecomp}B would align with the black dashed line of $|D_2,v=2\rangle$ throughout the whole simulation.}

\paragraph{The concept of sudden excitation}
This detailed analysis of a stationary photoexcitation reveals how the coupled electron-nuclear dynamics during photoexcitation is encoded in the BH representation. The discrete set of electronic states gives rise to the concept of sudden/vertical excitation, which reflects the fast initial response of the electrons to the laser field compared to the slower nuclear response. As such, the sudden excitation prompts us to picture photoexcitation as a vertical promotion of nuclei to a higher electronic state. The vertically promoted nuclear density then evolves on the given electronic potential to form the final state, connecting this final state smoothly to the initial state. 
The concept of sudden excitation becomes redundant once we focus on the total nuclear density as obtained in the EF representation. The total nuclear density must evolve smoothly and cannot suddenly change its position, as seen in the EF simulation. The initial response of the molecule to the electric field at the very beginning of the laser pulse is purely electronic, and the EF ansatz fully captures this in the time-dependent electronic wave function. The nuclei are impacted only later by responding to the electrons. Hence, the whole process can be understood without any vertical excitation, which belongs solely to the BH expansion, and without the use of a set of electronic states.

\paragraph{On the resonance condition}
Finally, we briefly comment on the concept of the resonance condition. The pulse spectral intensity (square of the pulse energy spectrum) reflects the energy composition of the complete laser pulse, i.e., of the pulse over its entire duration. Therefore, the resonance condition based on the pulse spectral intensity allows us to infer the final molecular state after the photoexcitation. Since our pulse spectral intensity encompasses solely the transition to the $|D_2,v=2\rangle$ state, only the resonant $|D_1,v=0\rangle$ and $|D_2,v=2\rangle$ states should be populated after the laser pulse interaction is over, which is consistent with our quantum dynamics simulations. 
However, the resonance condition based on the pulse spectral intensity does not apply to dynamics during photoexcitation, as demonstrated in Fig.~\ref{fig:trunkHspace}A. At the beginning of the laser pulse, the corresponding (transient) spectral intensity would be much wider in energy due to the time-energy uncertainty principle, targeting a broader range of transitions.\footnote{The short-time duration and wide energy spectrum justify the vertical excitation concept observed at the beginning of our dynamics.} 
As the pulse progresses, the energy spectrum narrows, these transitions are getting out of resonance, and populations of the corresponding states are decreasing.
As a thought experiment, abruptly (and artificially) stopping our laser pulse after its maximum at 50~fs would mean that all the neighbouring vibrational states depicted in Fig.~\ref{fig:trunkHspace}B are in resonance with the initial state.\footnote{We note that our discussion follows the populations of the off-resonant vibrational states of $D_1$ and $D_2$, completely omitting the off-resonant \textit{electronic} states $D_3$ and $D_4$. By the same mechanism, these off-resonant electronic states were also populated; however, they landed only 0.06\% population at their peak---a negligible number compared to 1.5\% of the vibrational $|D_2,v=1\rangle$ state.}
Therefore, considerations based on the pulse spectral intensity and resonance condition are valid for the molecular states of the system after photoexcitation, but cannot be relied on for describing dynamics during photoexcitation.

\subsection{Photoexcitation to a coherent superposition of electronic states}

In this Section, our attention will be focused on ultrashort, few-cycle laser pulses: the cornerstone of attochemistry. Such ultrashort pulses typically create a molecular state in a coherent superposition of electronically excited states; in the present case, the 1-fs pulse targets all the excited electronic states $D_2$ to $D_4$ of OH (see the spectral intensity in Fig.~\ref{fig:oh_ham}C). Photoexcitation with this 1-fs pulse will, therefore, also promote OH to its dissociative electronic states, unlike the 100-fs pulse. Hence, this ultrashort pulse allows us to investigate how attochemistry emerges from the BH and the EF representations.

\begin{figure}[ht!]
    \centering
    \includegraphics[width=1.0\linewidth]{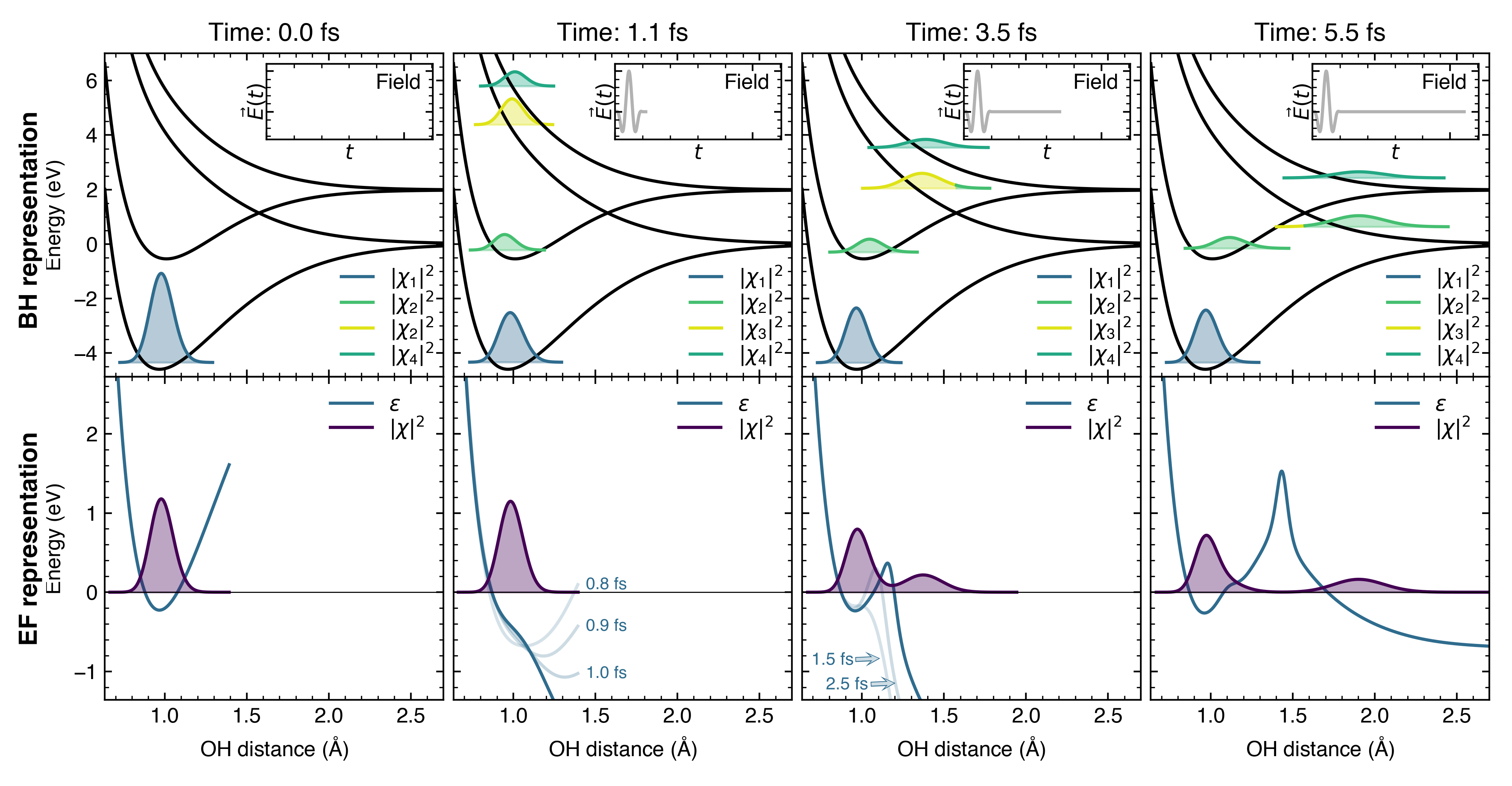}
    \caption{Snapshots of the photoexcitation of OH with a 1-fs laser pulse in the BH (top row) and the EF (bottom row) representations at four times: 0, 1.1, 3.5, and 5.5 fs. For the BH representation, we plot the BH nuclear densities for the adiabatic states ($|\chi_I|^2$) and their corresponding adiabatic potential energy curves (black). For the EF representation, we plot the total nuclear density ($|\chi|^2$) and the TDPES ($\tdpes$). The transparent lines in the bottom row depict TDPES at other snapshot times. The insets contain the laser pulse electric field $\vec{E}(t)$ up to the given time of the snapshot. An animated version of the plot is available using the following \href{https://github.com/JanosJiri/Photoexcitation-From-Different-Perspectives/blob/main/1-fs pulse/BH_vs_EF_A0.gif}{link}.\cite{movierepo} }
    \label{fig:attoA0}
\end{figure}

We begin the discussion with the BH representation again (top row of Fig.~\ref{fig:attoA0}). 
The laser pulse leads to nearly equal coherent populations of all three excited electronic states, transferring the nuclear amplitudes vertically between the electronic states without any visible nuclear motion (see the animated plot available using the following \href{https://github.com/JanosJiri/Photoexcitation-From-Different-Perspectives/blob/main/1-fs pulse/BH_vs_EF_A0.gif}{link}).\cite{movierepo} The excited-state densities right after the pulse (1.1 fs in the top row of Fig.~\ref{fig:attoA0}) resemble the ground-state nuclear density in both shape and position, in line with the concept of sudden/vertical excitation discussed in the previous Section. As time progresses, the nuclear densities in $D_3$ and $D_4$ follow their respective dissociative electronic curves, while the nuclear densities in $D_1$ and $D_2$ remain in the bound regions of their electronic potentials. Between 3.5 and 5.5~fs, the dissociating nuclear wave packet in $D_3$ passes the $D_3$/$D_2$ crossing point and transfers amplitude to the lower electronic state ($D_2$).

Let us now move to the EF representation for the very same photoexcitation process (bottom row of Fig.~\ref{fig:attoA0}). During the laser pulse, the originally bound TDPES (mimicking the bound $D_1$ electronic curve) experiences two separate motions: (\textit{i}) it oscillates up and down following the electric field of the pulse without changing its shape, and (\textit{ii}) it gets shallow towards larger bond distances near the end of the pulse (after around 0.8~fs), eventually opening a dissociation channel. Both motions are better seen in the animated version of the plot available using the following \href{https://github.com/JanosJiri/Photoexcitation-From-Different-Perspectives/blob/main/1-fs pulse/BH_vs_EF_A0.gif}{link}.\cite{movierepo} Once the laser pulse is over, the path to dissociation created by the TDPES is fully open and barrierless, and the nuclear density evolves towards larger OH bond distances: the molecule starts dissociating. At 3.5~fs, the nuclear density is almost split into two parts, now separated by a small barrier in the TDPES. At 5.5~fs, the two densities are completely separated, evolving on their corresponding parts of TDPES. These two separate minima of the TDPES are connected by a large barrier, causing the complete disconnection of the two nuclear densities.

Contrary to the long 100-fs pulse, the ultrashort 1-fs laser pulse does not trigger any significant nuclear dynamics during the photoexcitation; only the electrons respond within this time scale. How is this pulse scenario portrayed in the two representations? 
In the BH representation, this purely electronic response to the field appears as a vertical promotion of nuclear amplitudes, forming a so-called coherent superposition of electronic states. The actual nuclear dynamics sets in only after the pulse, when the same nuclear amplitudes begin to evolve on their corresponding potential energy surfaces. However, the nuclear dynamics are also affected by the electronic coherences, which impose an additional force on the nuclei\cite{Cardosa-Gutierrez2024}  yet are not directly seen in Fig.~\ref{fig:attoA0}.
Hence, the nuclear dynamics might be cumbersome to infer from the nuclear amplitudes.

The EF representation offers a different perspective on the nuclear motion resulting from a photoexcitation with an ultrashort pulse. The total nuclear density remains unchanged during the pulse, fully reflecting the absence of nuclear response to the ultrashort laser pulse. The nuclear dynamics begin just at the end of the pulse, when the nuclei start feeling the driving force created by the excited electrons, manifested in the TDPES. Thus, the EF representation offers a clear picture of the nuclear motion driven by the electrons via the TDPES.  

\section{Conclusions}
\label{sec:conclusions}

In this work, we examined the photoexcitation process in the OH radical using two different representations of the molecular wave function: the Born--Huang (BH) expansion and the exact factorization (EF), comparing their conceptual implications. We find that the EF representation offers a compelling picture for the formation of an excited molecular state upon photoexcitation: the EF clearly distinguishes nuclear and electronic responses to the laser field during photoexcitation. 
In contrast, the BH representation intertwines both electronic and nuclear responses to the field within the nuclear amplitudes, complicating their separate analysis. 

The four-state model of the OH molecule developed for this work allowed us to explore two distinct photoexcitation scenarios: irradiation of the molecule by a long pulse and an ultrashort pulse. In the long-pulse scenario, a 100-fs pulse targets a single excited vibronic state, forming a \textit{stationary} final molecular state. The dynamics connecting the initial and final stationary molecular states can thus be attributed purely to the photoexcitation process. 
Both representations depict this photoexcitation scenario using different concepts: the BH representation relies on the amplitude transfer between electronic states, while the EF employs barriers on the TDPES and tunneling to steer the nuclear dynamics triggered by the photoexcitation process. The picture of photoexcitation with tunneling offers a novel perspective on photoexcitation dynamics beyond BH amplitude transfer between electronic states.

The EF offered a somewhat clearer conceptual picture for investigating nuclear dynamics during photoexcitation: while the EF uses one compact nuclear wave function for all nuclei, the BH representation creates the notion of separate bundles of nuclei distributed among the electronic states.
In the present case, the EF nuclear density clearly depicts the nuclear motion by smoothly evolving between the initial and final states, guided by the TDPES.
On the contrary, the two BH (adiabatic) nuclear amplitudes involved might appear seemingly stationary, only transferring population between the resonant vibronic states. Uncovering the full nuclear dynamics---clearly observed in the EF---requires a closer investigation of the two nuclear amplitudes. The well-known concept of vertical excitation emerges from such an investigation, indicating that the initial nuclear amplitude promoted to the excited state is a projection of the ground-state amplitude. 
However, the vertical excitation concept only demonstrates that nuclei remain stationary during the excitation of the electrons, and is redundant in the EF framework. 

Furthermore, we touched upon the concept of the resonance condition and its use in photodynamics. The resonance condition dictates what states are populated after the laser pulse interaction, making it a useful tool for selecting initial states for the subsequent excited-state dynamics without modelling the photoexcitation explicitly. Excluding the off-resonant states in such cases is a common practice. However, we demonstrated that the off-resonant states are essential during photoexcitation: they facilitate the formation of the final resonant states, albeit they disappear once the interaction ceases. The neglect of off-resonant states can have a crucial impact on the final state formed by the pulse. An insufficient basis of (off-resonant) states in methods for explicit modelling of photoexcitation, such as XFAIMS\cite{Mignolet2016} or EOE\cite{Grunewald2025}, can have a quality-determining impact on the results.

In the ultrashort (attochemical) pulse regime, only the light electrons manage to respond to the few-cycle pulse. The BH representation depicts this event as a vertical excitation of nuclear amplitudes, creating a coherent superposition of electronic states---an electronic wave packet. The following evolution of nuclear amplitudes then captures both the dissociation of nuclei and damping of the electronic coherences, again blending electronic and nuclear dynamics. 
In the EF picture, the molecular response to the field is captured purely by the electronic wave function. The nuclear density remains untouched during the pulse and starts evolving only later in response to the excited electrons (via TDPES). Hence, the EF allows us to separate the fast electronic dynamics driven by attosecond pulses from the following nuclear dynamics driven by the excited electronic wave packet. 

This separation can be advantageous for practical attochemical simulations, often relying on static nuclei or mean-field Ehrenfest trajectories.\cite{Calegari2023, Lara-Astiaso2016, Lara-Astiaso2018,Palacios2020} Both approaches benefit from the techniques of time-dependent quantum chemistry for propagating the electronic dynamics coupled to laser pulses, incompatible with standard methods for nonadiabatic dynamics based on the BH expansion. However, they suffer from the lack of decoherence due to nuclear motions,\cite{Mannouch2024, Tran2024} which damps the electronic coherence on a short femtosecond timescale.\cite{Vacher2017, Calegari2023} 
Recent efforts to apply trajectory surface hopping methods, the workhorse of femtochemistry, for simulations with coherent superpositions of electronic states were also shipwrecked on the cliffs of decoherence,\cite{VillasecoArribas2024, Mannouch2024, Tran2024} highlighting the need for nonadiabatic methodologies suitable for attochemistry.\cite{Calegari2023} The presented results suggest that methods based on the EF, such as CT-MQC,\cite{Min2015} offer potential alternatives for simulating attochemistry.
CT-MQC possesses the same advantages as Ehrenfest dynamics, i.e., working with a single time-dependent electronic wave function, while capturing decoherence effects naturally, albeit approximately.\cite{Curchod2018b,Agostini2021} 

To summarize, our work highlights that the EF of the molecular wave function offers an alternative conceptual tool to the BH representation for describing photoexcitation in molecules. EF benefits from the clear separation of the nuclear and electronic responses to the field and the ensuing dynamics. This separation not only offers a different strategy to analyze nuclear dynamics during photoexcitation but also makes EF a potentially promising framework for attochemistry simulations, e.g., with its trajectory-based CT-MQC method.

\begin{acknowledgement}

Jiří Janoš thanks Jiří Suchan for insightful discussions on photoexcitation and the resonance condition. 
This work was supported by the grant of Specific university research - grant No A2\_FCHI\_2025\_053.
The authors thank the Czech Science Foundation for the support via grant number 26-22810S.
This work was supported by the project "The Energy Conversion and Storage", funded as project No. CZ.02.01.01/00/22\textunderscore 008/0004617 by Programme Johannes Amos Comenius, call Excellent Research.
BFEC acknowledges funding from EPSRC for the grants EP/V026690/1, EP/Y01930X/1, and EP/X026973/1.

\end{acknowledgement}

\begin{suppinfo}

The Supporting Information contains: details of fitting the OH radical model Hamiltonian, an extended set of quantum dynamics simulations with a truncated basis set, and a detailed decomposition of the EF quantities along the dynamics in different gauges. (PDF)

\end{suppinfo}

\providecommand{\latin}[1]{#1}
\makeatletter
\providecommand{\doi}
  {\begingroup\let\do\@makeother\dospecials
  \catcode`\{=1 \catcode`\}=2 \doi@aux}
\providecommand{\doi@aux}[1]{\endgroup\texttt{#1}}
\makeatother
\providecommand*\mcitethebibliography{\thebibliography}
\csname @ifundefined\endcsname{endmcitethebibliography}  {\let\endmcitethebibliography\endthebibliography}{}

\end{document}


\maketitle
\singlespacing
{ \hypersetup{hidelinks} \tableofcontents }

\clearpage
\section{Fitting the potential energy curves and transition dipole moments to build a four-state model Hamiltonian for the OH radical}

\begin{figure}[b!]
    \centering
    \includegraphics[width=1.0\linewidth]{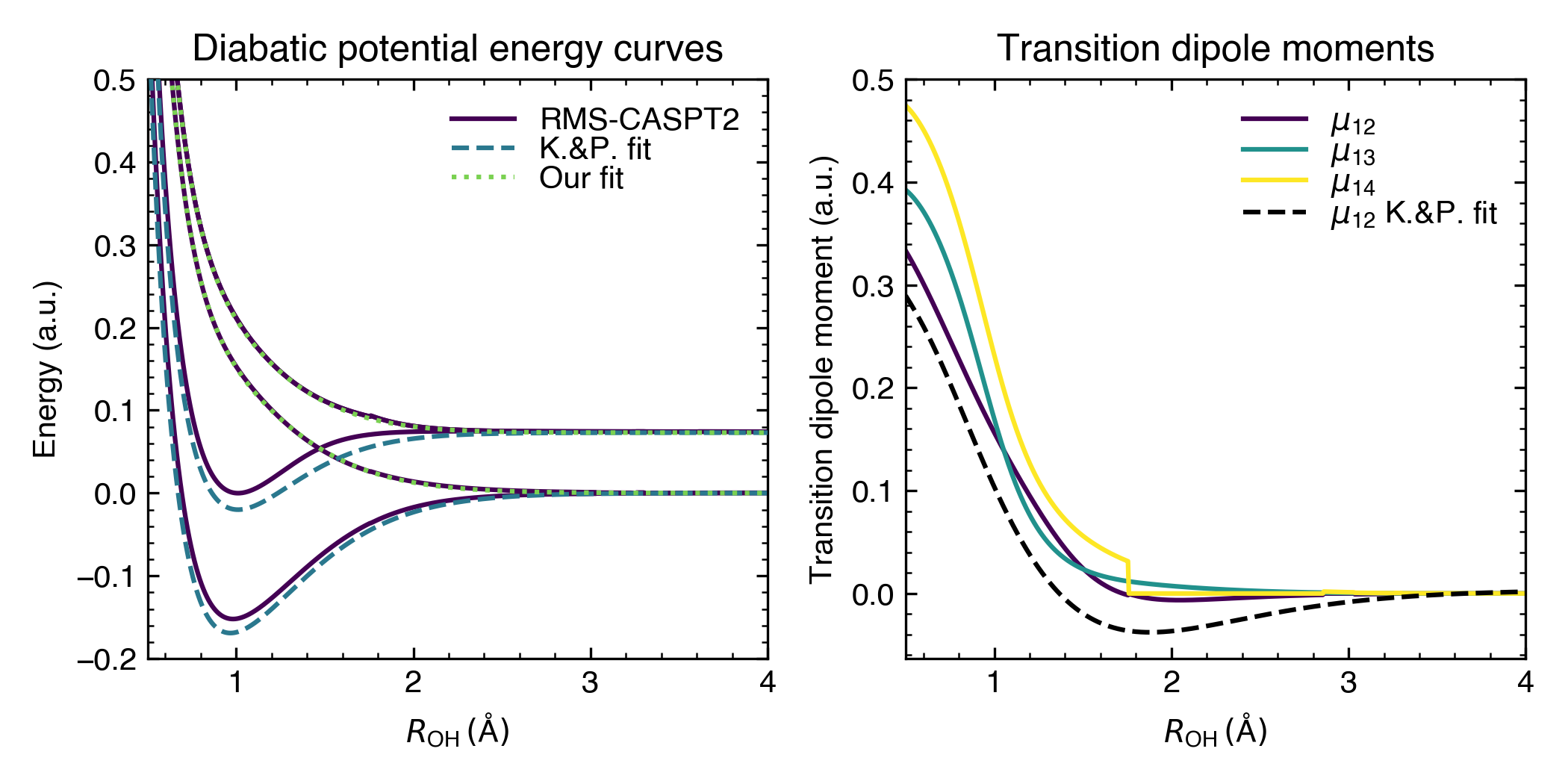}
    \caption{Left Panel: Diabatic potential energy curves of the OH radical calculated at the RMS-CASPT2 level of theory (solid violet lines). The fits by Korolkov and Paramonov\cite{Korolkov1998} (dashed blue lines) and our fit for the dissociative curves based on the RMS-CASPT2 potentials (dotted green lines) are included as well. Right Panel: Transition dipole moments calculated at the RMS-CASPT2 level of theory (solid) and the fit proposed by Korolkov and Paramonov\cite{Korolkov1998} (dashed) for these quantities.}
    \label{fig:ohfit}
\end{figure}

As stated in the main text, our model Hamiltonian for the hydroxyl radical is based on a Hamiltonian developed by Korolkov and Paramonov.\cite{Korolkov1998} Their diabatic Hamiltonian consists of two electronic states---$\mathrm{X}^2\Pi$ and $\mathrm{A}^2\Sigma^+$---which were fitted to extended Rydberg functions (see the main text for a definition). In this work, we extended this Hamiltonian by including two more electronic states: $\mathrm{1}^2\Sigma^-$ and $1^2\Delta$. For that purpose, we calculated the potential energy curves (PECs) of the OH radical at the RMS-CASPT2 level of theory with a cc-pVTZ basis for O and a cc-pVDZ for H in OpenMolcas version 22.06.\cite{openmolcas} The active space comprised 7 electrons in 5 orbitals (full valence active space) and was averaged over six electronic states. The calculated PECs are depicted in Fig.~\ref{fig:ohfit}. The dashed blue lines represent the original diabatic PECs of $\mathrm{X}^2\Pi$ and $\mathrm{A}^2\Sigma^+$ fitted by Korolkov and Paramonov\cite{Korolkov1998}, while the dotted green lines show our fit of the $\mathrm{1}^2\Sigma^-$ and $1^2\Delta$ states. Note that the original PECs by Korolkov and Paramonov do not perfectly match the RMS-CASPT2 curves, as their minima are lower in energy. Fitting all the states based on the RMS-CASPT2 curves would be the most consistent approach, yet we wished to keep the parametrization of the lowest two states by Korolkov and Paramonov\cite{Korolkov1998} since we built on top of their work by using pulses tuned for their PECs. Hence, we fitted only the two dissociative states and added them to the Hamiltonian designed by Korolkov and Paramonov. Although not fully consistent with the RMS-CASPT2 calculations, this description of the potential energy curves preserves the original parametrization, and we note that the PECs are adequately described in the dissociation limit. Once again, we emphasize that our four-state model is only intended to preserve the main features of the potential energy curves of the OH radical. 

Let us now discuss the transition dipole moments (TDMs). First, the TDMs between excited states are zero due to symmetry, since transitions between $1^2\Sigma^-$, $\mathrm{A}^2\Sigma^+$ and $1^2\Delta$ are forbidden. On the other hand, the transitions from the ground state $\mathrm{X}^2\Pi$ to the excited states $1^2\Sigma^-$, $\mathrm{A}^2\Sigma^+$ and $1^2\Delta$ are allowed by symmetry. 
Since the original model of Korolkov and Paramonov contains only the TDM to the lowest excited state, we obtained the additional TDMs from our RMS-CASPT2 calculations for our extended model. To extract the TDMs, we decided to set the polarization of the laser field along the $x$-axis and extracted only the $x$ component of the TDMs instead of their total magnitudes, see the right panel of Fig.~\ref{fig:ohfit}. Note that extracting the $y$ component would lead to the same result. We also summed up contributions from the degenerate electronic states. The discontinuities in the TDMs as a function of the \ce{OH} distance appear due to orbital rotations in the active space upon dissociation, which partially changes the character of the electronic states. These discontinuities also manifest in the PECs, yet they are not visible at the scale of the plot.

The RMS-CASPT2 calculations revealed that all TDMs between the ground electronic state and excited states share a similar magnitude and $R$-dependence, comparable to the previous results by Korolkov and Paramonov. Although the TDM fitted by Korolkov and Paramonov is lower in magnitude compared to the RMS-CASPT2 calculations, we decided to keep it in our model as the laser pulses we adopt from their work use this specific TDM. The RMS-CASPT2 calculation indicates that setting all the TDMs between the ground state and the other excited states to the same functional form should be a good approximation and preserve the qualitative picture of OH photoexcitation in the proposed model, even though it implies that the model cannot be regarded as quantitative anymore.

Finally, we provide below the Molcas input file used for generating the PECs.

\begin{lstlisting}[language=Python, style=style]
/*RMS-CASPT2 OH radical calculation */

&GATEWAY
  Coord=oh.xyz
  Basis=O.cc-PVTZ,H.cc-PVDZ
  Group=Nosym

&SEWARD  

&SCF
  UHF 
  Charge=0
  Spin=2 

&RASSCF
  nactel  =  7  0  0  
  inactive=  1  
  ras2  =  5    
  spin = 2
  ciroot  =  6  6  1
  
&CASPT2
  rmultistate=6  1  2  3  4  5  6
  MAXIter=50
  properties

&RASSI
 Heff  
 TDMN  
  1.0D-20   
\end{lstlisting}

\clearpage
\section{Quantum dynamics within a truncated basis of vibronic states}

\begin{figure}[ht!]
    \centering
    \includegraphics[width=0.64\linewidth]{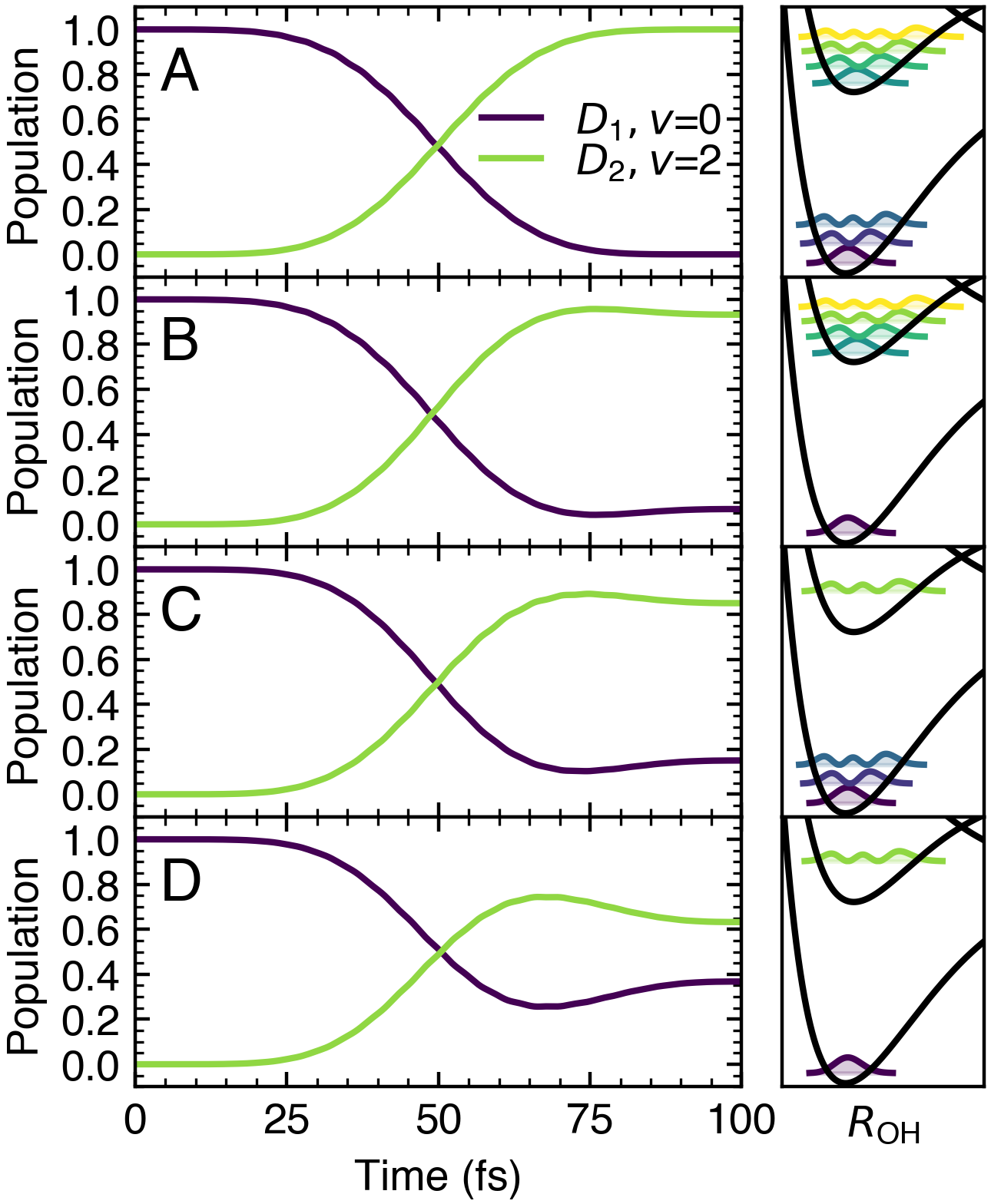}
    \caption{Populations of the $|D_1,v=0\rangle$ (purple) and $|D_2,v=2\rangle$ (green) states resulting from quantum-dynamics simulations with a truncated expansion of vibronic states: \textbf{A} seven vibronic states (three in $D_1$, four in $D_2$), \textbf{B} five vibronic states (one in $D_1$, four in $D_2$), \textbf{C} four vibronic states (three in $D_1$, one in $D_2$), \textbf{D} two vibronic states (one in $D_1$, one in $D_2$). The vibrational eigenstates used in the expansion are depicted in the right panels.}
    \label{fig:trunkHspace}
\end{figure}

We present additional quantum-dynamics simulations using a truncated basis of vibronic states and compared them to the results presented in the main text (Fig.~\ref{fig:trunkHspace}). Panels A (seven vibronic states facilitating full population transfer) and D (two-vibronic-state model with largely underestimated population transfer) are featured in the main text, and panels B and C will be discussed here. Fig.~\ref{fig:trunkHspace}B and Fig.~\ref{fig:trunkHspace}C show the effect of adding off-resonant vibrational states to only one of the electronic states considered. While both cases bring significant improvement over the simple two-vibronic-state model, neither of them results in a full population transfer. These results highlight that off-resonant vibrational states in both electronic states are necessary to fully capture the photoexcitation dynamics.

\clearpage
\section{Decomposition of the TDPES into its gauge-dependent and gauge-independent components, and test of different gauges}

\begin{figure}[ht!]
    \centering
    \includegraphics[width=0.94\linewidth]{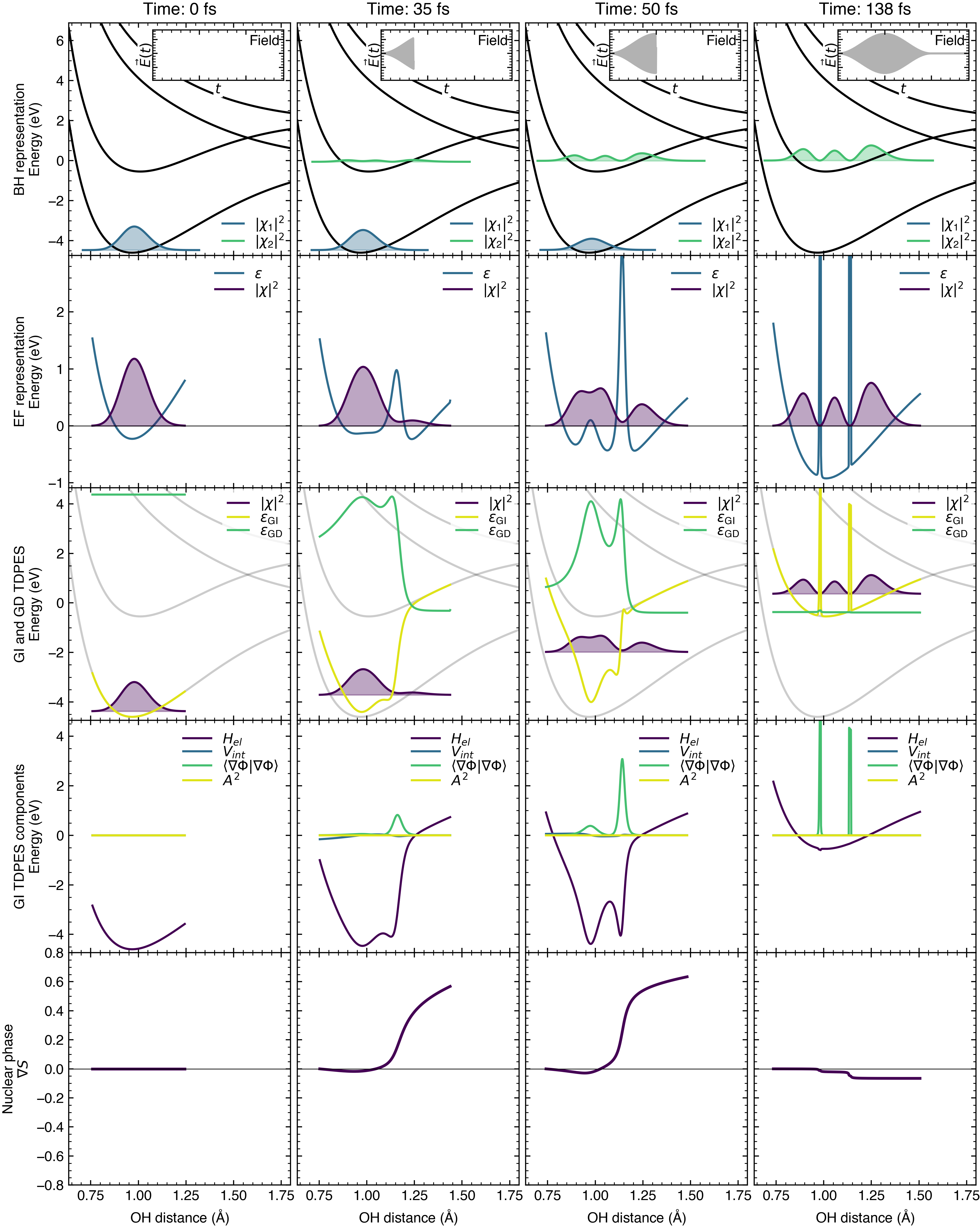}
    \caption{Decomposition of EF quantities for the photoexcitation of OH radical triggered by a 100-fs pulse in the zero-vector-potential gauge. Animation of the dynamics is available using the following \href{https://github.com/JanosJiri/Photoexcitation-From-Different-Perspectives/blob/main/100-fs pulse/BH_vs_EF_A0.gif}{link}.}
    \label{fig:100fsA0}
\end{figure}

\begin{figure}[ht!]
    \centering
    \includegraphics[width=0.94\linewidth]{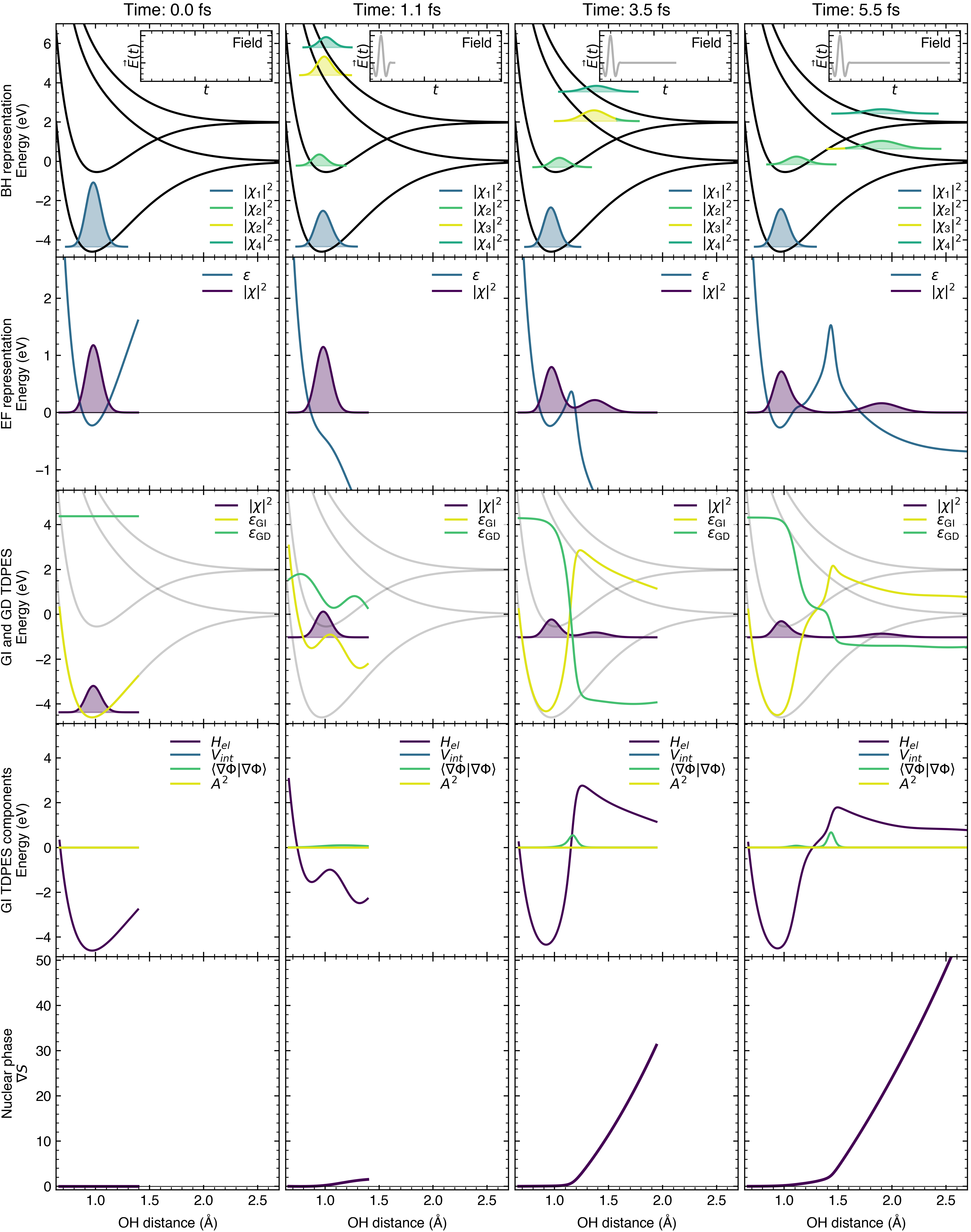}
    \caption{Decomposition of EF quantities for the photoexcitation of OH radical triggered by a 1-fs pulse in the zero-vector-potential gauge. Animation of the dynamics is available using the following \href{https://github.com/JanosJiri/Photoexcitation-From-Different-Perspectives/blob/main/1-fs pulse/BH_vs_EF_A0.gif}{link}.}
    \label{fig:1fsA0}
\end{figure}

\begin{figure}[ht!]
    \centering
    \includegraphics[width=0.94\linewidth]{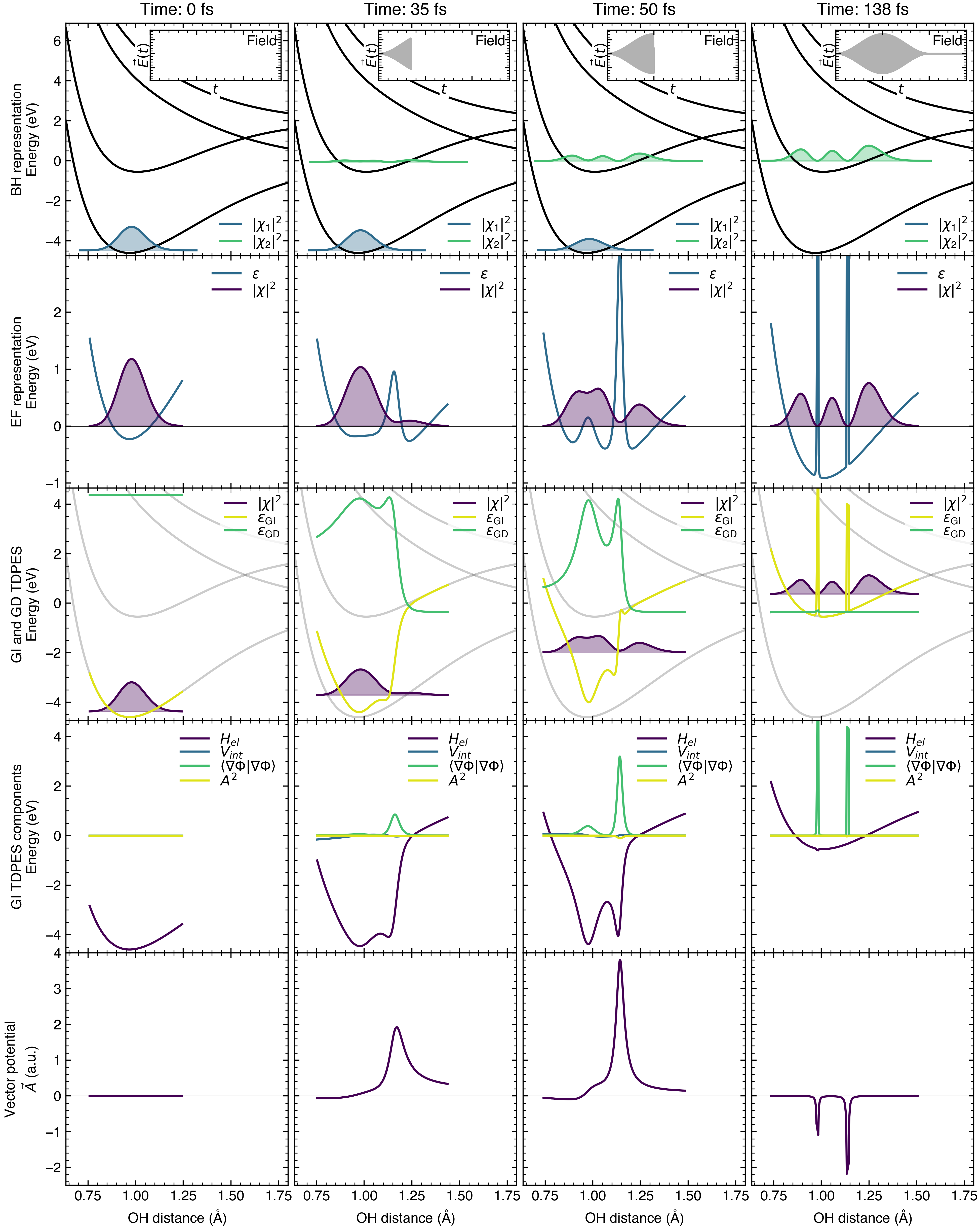}
    \caption{Decomposition of EF quantities for the photoexcitation of OH radical triggered by a 100-fs pulse in the zero-nuclear-phase gauge. Animation of the dynamics is available using the following \href{https://github.com/JanosJiri/Photoexcitation-From-Different-Perspectives/blob/main/100-fs pulse/BH_vs_EF_S0.gif}{link}.}
    \label{fig:100fsS0}
\end{figure}

\begin{figure}[ht!]
    \centering
    \includegraphics[width=0.94\linewidth]{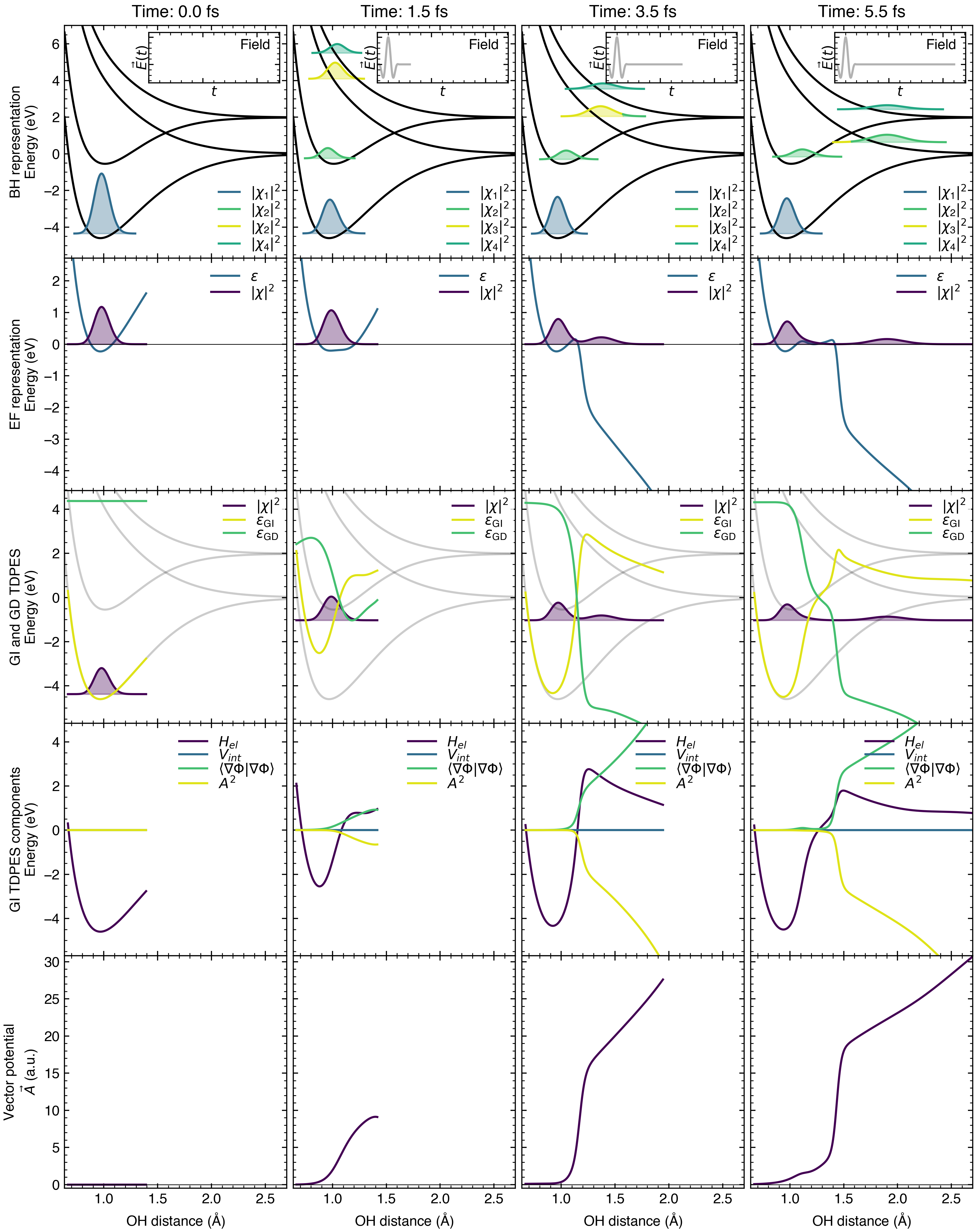}
    \caption{Decomposition of EF quantities for the photoexcitation of OH radical triggered by a 1-fs pulse in the zero-nuclear-phase gauge. Animation of the dynamics is available using the following \href{https://github.com/JanosJiri/Photoexcitation-From-Different-Perspectives/blob/main/1-fs pulse/BH_vs_EF_S0.gif}{link}.}
    \label{fig:1fsS0}
\end{figure}

We provide here a detailed analysis of the EF quantities during photoexcitation dynamics of the OH radical. The TDPES can be split into a gauge-dependent (GD) and a gauge-independent (GI) part as
\begin{equation}
    \tdpes(R,t) = \tdpesgi(R,t) + \tdpesgd(R,t) \,,
\end{equation}
where
\begin{align}
    \tdpesgd(R,t) &= - i \hbar \langle \Phi(R,t) | \frac{\partial}{\partial t} | \Phi(R,t) \rangle_{\rr} \,,\\
    \tdpesgi(R,t) &= \langle \Phi(R,t) | \hat{H}_\mathrm{el}  +\hat{V}_\mathrm{int}^e(\rr,t) | \Phi(R,t) \rangle_{\rr} + \frac{\langle \nabla \Phi(R,t)  |\nabla \Phi(R,t) \rangle_{\rr}}{2M} - \frac{A^2(R,t)}{2M} \,.
\end{align}
The EF dynamics presented in the main text for the 100-fs and 1-fs pulses used a gauge where the vector potential is set to zero [$A(R,t)=0$], and the components of the resulting TDPES are presented in Figs.~\ref{fig:100fsA0} and~\ref{fig:1fsA0}. The very same quantum dynamics is then expressed in the EF using a different gauge where the nuclear phase is now set to zero [$S(R,t)=0$], see Figs.~\ref{fig:100fsS0} and~\ref{fig:1fsS0}. Note that in this zero-nuclear-phase gauge, the nuclear dynamics are governed not only by TDPES but also by the vector potential. Hence, the dynamics cannot be interpreted using only TDPES, but also the vector potential, which represents the nuclear momentum field, must also be accounted for.

Note that for the dynamics resulting from a photoexcitation with the 100-fs pulse, both choices of gauge result in quantitatively similar dynamics for the TDPES. Hence, the gauge choice appears negligible in this case, and the same dynamics would be observed even for multidimensional cases where the zero-vector-potential gauge is not possible, and the zero-nuclear-phase gauge is typically used instead. The similarity of results from both gauges is not surprising, considering that the system features minimal momentum during the dynamics, and the vector potential is, therefore, negligible.
An animation for the quantum dynamics within the zero-nuclear-phase gauge is available using the following \href{https://github.com/JanosJiri/Photoexcitation-From-Different-Perspectives/blob/main/100-fs pulse/BH_vs_EF_S0.gif}{link}. For the dynamics associated to a photoexcitation with the 1-fs laser pulse, we observe that the TDPES behaves differently within the two gauges tested, especially when the laser pulse is over. At 5.5 fs and in the zero-nuclear-phase gauge (Fig.~\ref{fig:1fsS0}), the regions of TDPES supporting the split nuclear wave function are not separated by a barrier as in the zero-vector-potential gauge case, but by a step. The difference between the two gauges emerges from the momentum of the dissociating wave packet, which gives rise to a significant vector potential in the zero-nuclear-phase gauge. Hence, both TDPES and the vector potential are necessary for the interpretation of the dynamics. Animation for the zero-nuclear-phase gauge dynamics is available using the following \href{https://github.com/JanosJiri/Photoexcitation-From-Different-Perspectives/blob/main/1-fs pulse/BH_vs_EF_S0.gif}{link}.

\clearpage
\subsection{100-fs pulse in the zero-vector-potential gauge}

\clearpage
\subsection{1-fs pulse in the zero-vector-potential gauge}
\vspace{-0.1cm}

\clearpage
\subsection{100-fs pulse in the zero-nuclear-phase gauge}

\clearpage
\subsection{1-fs pulse in the zero-nuclear-phase gauge}

\clearpage
\providecommand{\latin}[1]{#1}
\makeatletter
\providecommand{\doi}
  {\begingroup\let\do\@makeother\dospecials
  \catcode`\{=1 \catcode`\}=2 \doi@aux}
\providecommand{\doi@aux}[1]{\endgroup\texttt{#1}}
\makeatother
\providecommand*\mcitethebibliography{\thebibliography}
\csname @ifundefined\endcsname{endmcitethebibliography}  {\let\endmcitethebibliography\endthebibliography}{}